\documentclass{svproc}
\usepackage{url}

\usepackage[normal,tight,center]{subfigure}
\usepackage{amssymb,amsmath,amsbsy}
\usepackage{graphicx}
\usepackage{epstopdf}
\usepackage{epsfig, psfrag}

\def\beq{\begin{equation}}
\def\eeq{\end{equation}}
\def\bea{\begin{eqnarray}}
\def\eea{\end{eqnarray}}
\def\mc{\mathcal}
\newcommand{\pd}{\partial}
\newcommand{\A}{{\cal{A}}}
\newcommand{\B}{{\cal{B}}}
\begin{document}
\mainmatter              % start of a contribution
\title{Compact Objects in EsGB Theory and beyond}
\titlerunning{Compact Objects in EsGB Theory and beyond}  % abbreviated title (for running head)
\author{Panagiota Kanti}
\authorrunning{Panagiota Kanti} % abbreviated author list (for running head)
\tocauthor{}
\institute{Theory Division, Physics Department, University of Ioannina, \\Ioannina GR 451 10, Greece\\
\email{pkanti@uoi.gr}
}

\maketitle              % typeset the title of the contribution

\begin{abstract}
In the context of General Relativity, black holes are not allowed to possess scalar hair, wormholes are not traversable and particle-like solutions are irregular. Therefore, in order to derive novel and physically interesting solutions that describe compact objects one needs to address generalised gravitational theories. One popular class of such theories is the Einstein-scalar-Gauss-Bonnet (EsGB) theory with a general coupling function between the scalar field of the theory and the quadratic Gauss-Bonnet term. Starting from black holes, we present a variety of spherically-symmetric solutions for several different forms of the coupling function and discuss their main features. We then proceed to wormhole solutions and demonstrate that the EsGB theory naturally supports traversable wormholes without the need for exotic matter. Regular scalarised particle-like solutions also emerge in the context of the same theory which also possess interesting observable features such as photon rings and echoes. Moving beyond this class of theories, we then address the more extended scalar-tensor Horndeski theory, briefly mention the types of black-hole solutions that arise, and demonstrate that an appropriately constructed disformal transformation of a black-hole solution, such as the Lu-Pang solution, results into a traversable wormhole in the context of the beyond-Horndeski theory. 
% We would like to encourage you to list your keywords within
% the abstract section using the \keywords{...} command.
\end{abstract}

\section{Introduction}

In 1915, Albert Einstein formulated the General Theory of Relativity \cite{Einstein}, a mathematically beautiful and,
at the same time, a physically relevant theory which has so far passed all experimental tests. It is, in addition, a much
more interesting theory compared to Newtonian Gravity since, not only does it describe the gravitational interactions
between two massive bodies, but it also incorporates the equivalence between mass and energy, and proceeds
further to predict the existence of new gravitational solutions, such as black holes or wormholes, and new phenomena
associated with them.

However, General Relativity (GR) is not a perfect theory -- if such a thing ever exists. On the cosmological side, the
Standard Cosmological Model, which has been formulated on the mathematical and physical basis provided by GR,
has a number of open problems: the nature of dark matter and dark energy, the coincidence problem, the spacetime
singularities, the right model for inflation, to mention a few. Also, the prospect of the unification of gravity with the
other forces seems unlikely within the GR -- the latter theory is a tensorial theory rather than a gauge field theory,
and is not renormalizable. But even on the gravitational side, GR is a rather restricted theory predicting in fact the
existence of a limited number of types of compact objects beyond stars, some of them with undesirable properties:
black holes obey ``no-hair'' theorems, wormholes cannot keep their tunnels open or are plagued by internal singularities
and gravitational particle-like solutions (solitons) simply do not exist.

Therefore, in order to find new black-hole solutions, traversable wormholes or particle-like solutions,  going  beyond
General Relativity seems to be a one-way road. For this reason, a large number of generalised theories of gravity
have been formulated by adding to the Einstein-Hilbert action new fields, new gravitational terms and couplings
among them.  The simplest extension of GR is the scalar-tensor theories \cite{Bronnikov}-\cite{Charmousis_rev},
however, as it quite well known, one must consider non-minimally coupled scalar fields in order for
physically-interesting solutions to emerge. 

In this review, we will focus first on the Einstein-scalar-Gauss-Bonnet (EsGB) theory as an indicative example of
a non-minimal-coupled scalar-tensor theory. It is a rather simple theory containing a single scalar degree of freedom
coupled to the quadratic, gravitational Gauss-Bonnet term. This theory has been studied for decades, as it arises in
the context of superstring effective theory at low energies \cite{Callan}-\cite{Metsaev} or in Kaluza-Klein compactifications of
Lovelock's theory \cite{Lovelock}. Due to the presence of the quadratic Gauss-Bonnet (GB) term, this generalised
theory reduces to GR, with a trivial scalar field, in the limit of weak gravity but may lead to important modifications
from GR in the strong-gravity regime. These modifications include novel gravitational solutions characterised by a
non-trivial scalar field. Thus, in the context of the EsGB theory, we will search for new scalarised solutions describing
various types of compact objects. As we will demonstrate, the configuration space of solutions of the EsGB
theory is indeed a particularly rich one, with some areas of it still waiting to be explored.

The EsGB theory is in fact a subclass of Horndeski theory \cite{Horndeski}, which is a the most general scalar-tensor
theory with field equations containing only up to 2nd order derivatives of the metric tensor and the scalar field. 
The scalar field is now allowed to have a number of couplings with different gravitational terms and derivative-like
couplings as well. The theory can be further generalised to the beyond-Horndeski theory via the addition of
two additional terms in the Lagrangian. In the second part of this review, we will focus on the beyond-Horndeski
theory, and demonstrate that, in its context, new solutions describing black holes with an (Anti-)de Sitter-Reissner-Nordstrom
asymptotic behaviour at large distances may be analytically derived. For the derivation of wormhole solutions,
we will employ an alternative method, that of applying a disformal transformation to a known solution
of the theory.  We will show that such an approach can easily lead to new wormhole solutions with a number
of particularly attractive properties. 

The outline of this review is as follows: in Section 2, we will give a brief overview of the type of compact objects arising
in the context of GR and of the minimal Einstein-scalar theory. In Section 3, we will present the EsGB theory, and
search for black holes, wormholes and particle-like solutions in its context. In Section 4, we will proceed to consider
the beyond-Horndeski theory, and present some recent results on black holes and wormhole solutions. We will
present our conclusions and some food-for-thought comments for the future in Section 5.

%%%%%%%%%%%%%%%%%%%%%%%%%%%%%%%%%%%%%%%%%%%
%\section{ }
\section{Compact Objects in GR and Einstein-scalar theory}

In this section, we will briefly discuss the main characteristics of three types of compact objects that
will be of interest to us, namely black holes, wormholes and particle-like solutions, as these arise in the
context of General Relativity and Einstein-scalar theory. 

\subsection{Black Holes}

As is well know, General Relativity admits only three families of black-hole solutions: the Schwarzschild
solution \cite{Schwarz}, which describes a spherically-symmetric, neutral black hole, the Reissner-Nordstrom
solution \cite{Reissner}, which also describes a spherically-symmetric but charged black hole,
and the Kerr(-Newman) solution \cite{Kerr}, which describes a rotating, neutral (or, charged) black hole. 
According to the ``no-hair'' theorems of GR \cite{NH-scalar}, a BH may be characterized at most by only
three parameters -- three physical, conserved quantities -- namely its mass $M$, electromagnetic charge
$Q$ and angular-momentum $J$. All three aforementioned GR solutions perfectly comply with this dictation. 

The simplest extension of pure General Relativity amounts to the addition of a free, massless scalar field
$\phi$ to the theory. If we assume that this field is a static, spherically-symmetric one, i.e. $\phi=\phi(r)$, 
its equation of motion simply reads
\begin{equation}
\Box \,\phi=0 \,\Rightarrow \, \partial_r [\sqrt{-g} \,g^{rr}\,\partial_r \phi] =0\,.
\end{equation}
For an invertible metric tensor with $g \neq 0$, the above leads to the result
\begin{equation}
\phi' \sim g_{rr}\,,
\end{equation}
which diverges at the horizon. As a result, the simple Einstein-scalar theory does not possess a regular,
scalarised black-hole solution.

\medskip
Even for a self-interacting scalar field with $V(\phi) \neq 0$, Bekenstein's  {\it old scalar ``no-hair theorem''}
\cite{NH-scalar} excludes regular black-hole solutions in a general class of minimally-coupled scalar-tensor
theories. Starting from the scalar-field equation, multiplying with $\phi$, integrating over the exterior spacetime,
and performing an integration by parts of the $\Box \phi$ term, we obtain \cite{Herdeiro-review}
\begin{equation}
\int_{\cal V} d^4 x \,\sqrt{-g} \left[ \partial_\mu \phi \,\partial^\mu \phi +\phi V'(\phi)\right] + 
\int_{\cal \partial V} d^3x \sqrt{h} \,\eta^\mu \,\phi \partial_\mu \phi =0\,.
\end{equation}
The boundary term at the end of the above expression vanishes both at the horizon of the black hole, provided
that the scalar field remains there finite, and at asymptotic infinity, under the assumption that the self-interacting scalar field
falls off sufficiently fast.  The first term inside brackets gives $g^{rr} (\partial_r \phi)^2>0$ and, thus, the constraint
allows for scalarised black-hole solutions only in the case where $\phi V'(\phi) <0$. However, for e.g. a typical
mass term $V(\phi)=m^2 \phi^2/2$, this would demand $m^2<0$ and would result in a non-physical theory. 
Generalising this, we may therefore conclude that any theory with a minimally-coupled scalar field and a
self-interacting potential
satisfying the constraint $\phi V'(\phi)>0$ does not allow for black-hole solutions with a non-trivial, static scalar hair.

%%%%%%%%%%%%%%%%%%%%%%%%%%%%%%%%%%%%%%%%%%%
%\section{ }
\subsection{Wormholes} 

General Relativity admits a second class of compact objects, namely wormholes, which are in fact hidden
in the interior of all black-hole solutions that the theory predicts. Unfortunately, these wormhole solutions
are not {\it traversable}. 

Taking the Schwarzschild black-hole line-element as a paradigm
%%%%%%
\begin{equation}
ds^2 = - \left(1-\frac{2M}{r}\right)\,dt^2 + \left(1-\frac{2M}{r}\right)^{-1}\,dr^2 + 
r^2\,(d\theta^2 + \sin^2 \theta\,d \varphi^2)\,,
\end{equation}
it is clear that whereas the exterior region ($r>2M$) of this background is clearly static, the interior region
($r<2M$), due to the change of its signature as we cross the horizon, is clearly dynamical. Even the presence of the spacetime
singularity at $r=0$ is not in fact a static feature of the complete Schwarzschild spacetime.  A simple
but careful analysis \cite{Misner} reveals that, as the time goes by, a throat appears in the place of the
singularity, which connects two asymptotically far-away regions. The radius of the throat expands,
reaches its maximum value $r_{max}=2M$ and then shrinks again and disappears leaving behind the
two spacetime singularities of the black and the white hole which together comprise the complete
Schwarzschild geometry. Unfortunately, this Einstein-Rosen passage \cite{Rosen}\cite{Wheeler}\cite{Visser}
opens and closes so quickly that no physical particles, including photons, can pass through.

The Reissner-Nordstrom and Kerr geometries also possess similar internal tunnels. In fact, due to the
presence of the internal Cauchy horizons in both of these solutions, the spacetime singularity can always
be avoided and the tunnel remains always open. Unfortunately, it is again non-traversable: the 
internal Cauchy horizons they both possess are unstable, and any small disturbance causes them to
collapse and turns them to a spacetime singularity, which unavoidably blocks the passage.

Trying to construct a more general theory than GR, in the context of which a traversable wormhole solution
could emerge, we may add as before a scalar field, free of self-interacting. One could also adopt a
different perspective \cite{MT} on how the spacetime should look like in order to avoid altogether the
presence of a horizon or a spacetime singularity, thus enhancing the probability for constructing a
traversable wormhole. As we will see in detail in Section 4, where we will also follow this different perspective,
the addition of a massless scalar field can indeed support a wormhole solution, the well-known Ellis-Bronnikov
wormhole \cite{Ellis}\cite{Bronnikov_worm}. However, the scalar field must be a {\it ghost} one since its energy density
$\rho$ satisfies the constraint $\rho<0$, and thus violates the energy conditions. The addition of a
self-interacting potential also leads to constraints on the energy density and pressure components of the
theory which again hint to some exotic form of matter rather than a physical field.

%%%%%%%%%%%%  SLIDE 16 %%%%%%%%%%%%%%%%%%%%
%\section{}
\subsection{ Particle-like solutions} 

In flat Minkowski spacetime, solutions that are regular and describe different types of distribution of
matter are quite common, and are usually termed {\it solitons}. However, in the context of a pure gravitational 
theory, such as General Relativity, no such regular solutions emerge.

The same conclusion holds when simple modifications of GR are considered. For instance, if we add
again a massless, spherically-symmetric scalar field to the theory, the well-known Fisher/Janis-Newman-Winicour-Wyman
solution \cite{Fisher}-\cite{Wyman} can be found where the metric components and scalar field are given by the expressions
\begin{equation}
|g_{tt}| \sim (r-r_s)^{2s}\,, \quad g_{rr} \sim (r-r_s)^{2(1-s)}\,, \quad \phi \sim D \ln  \left(r-r_s\right)\,,
\end{equation}
respectively, with $s=1/\sqrt{1+(D/2M)^2}$. In the above, $M$ is the mass of the solution and $D$ a scalar
``charge''. However, as one may see by calculating the gravitational scalar quantities
of the spacetime, this is an irregular solution since the latter diverge at the radius $r_s=M/2s$.  Similar
results follow if one adds a self-interacting potential for the scalar field.

%%%%%%%%%%%%  SLIDE 7 %%%%%%%%%%%%%%%%%%%%
\section{The Einstein-Scalar-GB Theory}

According to the brief review of compact objects presented in Section 2, one would need to consider a 
theory beyond pure GR as well as beyond the simple Einstein-scalar theory in order to discover novel solutions
describing compact objects which are physically interesting. A more elaborate, generalised theory of gravity
could follow by introducing extra fields and/or higher gravitational terms, and could be schematically described
by the following  action functional 
%%%%%%%%%%%%%%%%%
\begin{equation}
S=\int d^4 x\,\sqrt{-g}\,\Bigl[f(R, R_{\mu\nu}, R_{\mu\nu\rho\sigma}, \Phi_i)+ 
{\cal L}_X (\Phi_i)\Bigr]\,,
\end{equation}
where $R_{\mu\nu\rho\sigma}$ is the Riemann tensor, $R_{\mu\nu}$ the Ricci tensor, $R$ the Ricci scalar and
$\Phi_i$ stands collectively for the different types of fields present in the theory whose properties are described
by the Lagrangian ${\cal L}_X (\Phi_i)$. 
In what follows, we will retain,  apart from the gravitational field, a single additional, scalar degree of freedom
and ignore all other forms of fields. We will nevertheless introduce a higher-derivative term in the form of a
quadratic gravitational term. Such a term would naturally go unnoticed in regions of weak gravitational
field but could cause significant deviations from GR in the strong field regime. 

In particular, we will consider the following generalised, quadratic theory of gravity 
\begin{equation}
S=\int d^4 x\,\sqrt{-g}\,\left[\frac{R}{16 \pi G}-
\frac{1}{2}\,\partial_\mu \phi\,\partial^\mu \phi + 
f(\phi)\,R^2_{GB}\right], \label{action}
\end{equation}
%%%%%%%%%%%%%%%%
where $R^2_{GB}$ is the so-called Gauss-Bonnet (GB) term
%%%%%%%%%%%%%
\begin{equation}
R^2_{GB}=R_{\mu\nu\rho\sigma} R^{\mu\nu\rho\sigma}-
4 R_{\mu\nu} R^{\mu\nu} +R^2
\label{def_GB}
\end{equation}
%%%%%%%%
and $f(\phi)$ is an arbitrary coupling function between the scalar field $\phi$ and the GB term. 
The above theory is hardly a new one -- in fact, it is common knowledge that it arises as part of
the string effective action at low energies, as part of a Lovelock effective theory in four dimensions
or as part of an extended scalar-tensor (Horndeski or DHOST) theory. It is a higher-derivative,
gravitational theory of gravity yet simpler than one would expect: the particular combination of
the gravitational quantities appearing in the definition of the GB term (\ref{def_GB}) guarantees 
that the field equations contain only up to 2nd-order derivatives of the metric tensor and the
scalar field thus avoiding any Ostrogradski instabilities \cite{Ostrogradski}. It is in the context of this quadratic theory --
upgraded to a class of theories due to the general form of the coupling function $f(\phi)$ -- that we
will look for scalarised solutions describing novel black-holes, traversable wormholes and regular
particle-like solutions.

%%%%%%%%%%%%  SLIDE 9 %%%%%%%%%%%%%%%%%%%%
\subsection{Black-Hole Solutions in Einstein-Scalar-GB Theory}

Do we have any reason to believe that the generalised gravitational theory (\ref{action}) may
lead to scalarised black holes, in other words that this theory violates Bekenstein's scalar no-hair
theorem? Yes, quite a few indeed. To start with, more than 25 years ago, a novel type of black
holes, the so-called {\it dilatonic} black holes \cite{DBH1}\cite{DBH2}, were discovered in the context
of the theory  (\ref{action}) with an exponential coupling function, i.e. $f(\phi)=\alpha e^{\phi}$,
between the scalar field, or dilaton, and the GB term. Due to the presence of the GB term, the
field equations are quite complicated and cannot be solved analytically - therefore, the dilatonic
black holes, as well as a large number of variants of this solution were found either in an approximate 
orm or through numerical integration (see \cite{Gibbons}-\cite{Blazquez} for an indicative list of works). 

Two decades later, another class of scalarised black-hole solutions was found \cite{SZ}\cite{Benkel}
in the context of the shift-symmetric EsGB theory, i.e. for the choice $f(\phi)=\alpha \phi$
of the coupling function. Here, as in the case of the dilatonic black holes, too, $\alpha$ is a coupling
constant of the theory. Both the dilatonic and the scalarised shift-symmetric black-hole solutions
violated the requirements of the scalar no-hair theorems, both the old \cite{NH-scalar} and
the new versions \cite{Bekenstein_new}\cite{Sotiriou_no_hair}\cite{Nicolis_no_hair} of it, thus paving the way
for new black-hole solutions with characteristics not allowed by General Relativity
\cite{Herdeiro_Radu}\cite{Babichev_Charmousis}.

The obvious question which readily emerges is whether we can find additional classes of black-hole
solutions in the context of the theory (\ref{action}) but for other choices of the coupling function apart
from the exponential and the linear one. To investigate this, we reconsidered the Einstein-scalar-GB
theory but allowed for a general form of the coupling function $f(\phi)$ \cite{ABK1}\cite{ABK2}. We focused on
the simplest possible background, that of a static, spherically-symmetric black hole, and assumed
the following form of line-element
%%%%%%%%%%%%%%%%
\begin{equation}
{ds}^2=-e^{A(r)}{dt}^2+e^{B(r)}{dr}^2+r^2({d\theta}^2+\sin^2\theta\,{d\varphi}^2)\,,
\end{equation}
%%%%%%%%%%
with two unknown metric functions of the radial coordinate. Taking the variation of the action (\ref{action})
with respect to the  scalar field $\phi$ and the metric tensor $g_{\mu\nu}$, we obtain the scalar field and
gravitational field equations
\begin{equation}
\nabla^2 \phi+\dot{f}(\phi)R^2_{GB}=0\,, \qquad 
R_{\mu\nu}-\frac{1}{2}\,g_{\mu\nu}\,R=T_{\mu\nu}\,,
\label{set_field}
\end{equation}
respectively. In the above
\begin{equation}
T_{\mu\nu}=-\frac{1}{4}g_{\mu\nu}(\partial \phi)^2+
\frac{1}{2}\partial_{\mu}\phi\partial_{\nu}\phi -\frac{1}{2}(g_{\rho\mu}g_{\lambda\nu}+g_{\lambda\mu}g_{\rho\nu})
\eta^{\kappa\lambda\alpha\beta}\tilde{R}^{\rho\gamma}_{\quad\alpha\beta}
\nabla_{\gamma}\partial_{\kappa}f
\end{equation}
is the energy-momentum tensor of the theory which receives contributions from both the kinetic term of the scalar
field and its non-minimal coupling to the quadratic GB term.

The set of field equations (\ref{set_field}) involve the two unknown metric functions $A(r)$ and $B(r)$, and the
scalar field $\phi(r)$. In fact, the $(rr)$-component of the gravitational equations takes the form of a 2nd order
polynomial for $e^{B}$, which therefore may be determined once the solutions for $A(r)$ and $\phi(r)$ are
found. The remaining equations reduce indeed to a system of only two independent, ordinary differential
equations of second order for the unknown functions $A(r)$ and $\phi(r)$, which is given below schematically
\begin{equation}
A''=\frac{P}{S}\,,\qquad \phi''=\frac{Q}{S}\,. \label{fin_sys}
\end{equation}
The quantities $(P,Q,S)$ depend on $(r, \phi, \phi', A')$, and the interested reader may find their exact
expressions as well as the remaining technical details of this analysis in \cite{ABK1}\cite{ABK2}. 

Due to the complexity of the quantities involved, the integration of the system demands numerical integration. 
However, it may be solved analytically in the small and large regimes of the radial coordinate. The corresponding
solutions may be used as boundary solutions for the general solution of the system and, in addition, their form
will reveal under which constraints these describe indeed a robust black-hole background. 

Starting from the small $r$-regime, we demand the presence of the most important feature of black-hole
geometry, that of a regular horizon. For this, we impose the asymptotic behaviour 
\begin{equation}
e^{A(r)} \rightarrow 0, \quad e^{-B(r)} \rightarrow 0, \quad \phi(r) \rightarrow \phi_h\,,
\label{asym_rh}
\end{equation}
as $r$ approaches a certain value $r_h$. The last demand -- the finiteness of the scalar field -- encompasses
the notion of the regularity of the black-hole horizon.  The same must naturally hold for the first and second
derivative of the field.  However, demanding that, under the behaviour of the metric functions displayed
in (\ref{asym_rh}), $\phi''$ remains finite at the horizon $r_h$, the second of the equations in (\ref{fin_sys})
leads to the constraint
\begin{equation}
\phi'_h=\frac{r_h}{4\dot{f}_h}\left(-1\pm\sqrt{1-\frac{96\dot{f}_h^2}{r_h^4}}\right).
\label{phi'_con}
\end{equation}
The above formula ensures that, for a selected coupling function and value of the scalar field $\phi_h$ at the
black-hole horizon $r_h$, choosing this particular value for $\phi'_h$ leads to a gravitational background that
describes the near-horizon geometry of a black hole.  No constraint arises on the form of the coupling
function $f(\phi)$ itself, which therefore remains arbitrary. The coupling function needs to satisfy an
additional constraint which follows from the positivity of the expression under the square root in (\ref{phi'_con});
this may be written as 
\begin{equation}
\dot{f}_h^2<\frac{r_h^4}{96}\,,
\label{con_f}
\end{equation}
%%%%%%%%
and interpreted as a bound on the lower value of the black-hole horizon radius $r_h$ for a given $f(\phi)$ and
$\phi_h$. It is this characteristic that distinguishes the GB black hole from the Schwarzschild black hole, its
analog in the context of GR. Using the above results, the field equations may give the asymptotic forms of
the metric functions and scalar field near the horizon, which are found to have the form
\begin{equation}
e^{A}=a_1 (r-r_h) + ... \,, \qquad e^{-B}=b_1 (r-r_h) + ...\,,
\end{equation}
\begin{equation}
\phi =\phi_h + \phi_h'(r-r_h)+ \phi_h'' (r-r_h)^2+ ...\,,. \label{sol_rh}
\end{equation}

At the other asymptotic regime, i.e. at large distances from the horizon, we assume, as usually, a
power series expansion in $1/r$ for the three unknown functions, namely
%%%%%%%%%%%%%%%%%%
\beq
e^{A}=1+\sum_{n=1}^\infty{\frac{p_n}{r}}\,,\qquad 
e^{B}=1+\sum_{n=1}^\infty{\frac{q_n}{r}}\,,\qquad
\phi =\phi_{\infty}+\sum_{n=1}^{\infty}{\frac{d_n}{r}}\,.
\label{far-gen}
\eeq
%%%%%%%%%%%%%%%%%%
Substituting these expressions into the field equations, we may determine the arbitrary coefficients
$(p_n, q_n, d_n)$. In fact, $p_1$ and $d_1$ remain arbitrary and are identified with the black-hole mass
$M$ and scalar charge $D$, respectively. The asymptotic form of the metric and scalar field then take
the final form 
\begin{equation} e^A=\; 1-\frac{2M}{r}+\frac{MD^2}{12r^3}+..., \qquad 
e^B=\; 1+\frac{2M}{r}+\frac{16M^2-D^2}{4r^2}+...\,,
\end{equation}
\begin{equation}
\phi=\; \phi_{\infty}+\frac{D}{r}+\frac{MD}{r^2}+\frac{32M^2D-D^3}{24r^3}
+\frac{12M^3D-24M^2\dot{f}-MD^3}{6r^4}+...\,. \label{sol_far}
\end{equation}
As in the near-horizon analysis, no constraint on $f(\phi)$ arises by demanding a robust black-hole geometry
at large distances. A general coupling function $f$ does not interfere with the existence of an asymptotically-flat
limit with its exact form making an appearance not earlier than in the 4th order term of the expansion.

In order to construct a complete black-hole solution, we need to find a solution of the field equations
(\ref{fin_sys}) which smoothly interpolates between the two asymptotic forms (\ref{sol_rh}) and (\ref{sol_far}).
And this is where the no-hair theorems play a crucial role: if the theory (\ref{action}) satisfies the requirements
of the no-hair theorems, these dictate that no solution that smoothly connects these two asymptotic forms can be found. 
Let us look briefly into this. We will start from the old no-hair theorem and follow a similar analysis \cite{NH-scalar}
\cite{Herdeiro-review}. We will employ the scalar-field equation, multiply by $f(\phi)$ in our case, and integrate over the
entire exterior spacetime. Performing an integration by parts, we arrive at the constraint \cite{ABK1}\cite{ABK2}\cite{Alex}
\begin{equation}
\int_{\cal V} d^4 x \,\sqrt{-g} \,\dot f(\phi) \left[ \partial_\mu \phi\,\partial^\mu \phi-
 f(\phi)\,R^2_{GB}\right] - \int_{\cal \partial V} d^3x \sqrt{h} \,\eta^\mu f(\phi) \partial_\mu \phi =0\,.
\end{equation}
Due to the dependence of the scalar field solely on the radial coordinate, it holds that 
$\partial_\mu \phi\,\partial^\mu \phi=g^{rr} (\partial_r \phi)^2>0$ and $\eta^\mu \partial_\mu \phi=g^{rr}\partial_r \phi$.
The whole boundary term at the end of the above expression vanishes as usual at the black-hole horizon. However,
as noted in \cite{Alex}, its value at asymptotic infinity depends on the form of the coupling function.
If $f(\phi_\infty) =0$, then the boundary term vanishes altogether and we need to address only the
expression inside the square brackets in the first integral. One may easily see, by using the asymptotic
expressions (\ref{sol_rh}) and (\ref{sol_far}), that the GB term takes on positive values at the two
asymptotic regimes \cite{ABK1}\cite{ABK2} -- as the exact numerical analysis reveals, in fact it remains
positive over the entire exterior regime. Therefore, the emergence of black-hole solutions in this
case is allowed only for $f(\phi)>0$ according to the old no-hair theorem.  If, on the other hand,
$f(\phi_\infty) \neq 0$, then the boundary term at asymptotic infinity reduces to $f(\phi_\infty)  D$.
In this case, solutions arise again for $f(\phi)>0$ but also for a certain interval for negative values
of $f(\phi)$ \cite{Alex}.

Let us now turn to Bekenstein's new version of no-hair theorem \cite{Bekenstein_new}, developed
for a theory with a minimally-coupled scalar field, and examine
in turn whether the theory (\ref{action}) evades its own requirements, too. This theorem relies on the
particular profile of the $T^r_{\;\,r}$ component of the energy-momentum tensor. For instance, it demands
that, at asymptotic infinity, $T^r_{\;\,r}$  is positive and decreasing. Indeed, using the asymptotic
solution (\ref{sol_far}), and the exact expression of $T^r_{\;\,r}$, we find that this is indeed the case.
Our result therefore agrees with the one derived in \cite{Bekenstein_new} at large distances; this was anticipated
since the quadratic GB term is not expected to play any role far away from the black-hole horizon where the
curvature of spacetime is small. In the near-horizon regime, the novel no-hair theorem dictates that
$T^r_{\;\,r}$ is negative and increasing. In the context of the Einstein-scalar-GB theory, though, this
does not hold any more: indeed, if we use the near-horizon asymptotic solution (\ref{sol_rh}),  we find that 
%%%%%%%%%%%%%
\beq
sign(T^r_r)_h = -sign(\dot f_h \phi_h') = 1 \mp \sqrt{1-96 \dot f^2/r_h^4)}\,,
\eeq
%%%%%%%%%%
where we have used the constraint Eq. (\ref{phi'_con}) for the regularity of the horizon. It is clear that the
above expression is always positive-definite, in contrast to the requirement of the novel no-hair theorem.
The presence of the GB term near the horizon -- where the curvature is strong -- changes the profile of
$T^r_{\;\,r}$, and causes the evasion of Bekenstein's new no-hair theorem. 

\begin{figure}[t]
%\hspace*{5.0cm}
\mbox{\hspace*{-0.0cm}\includegraphics[height=1.6in]{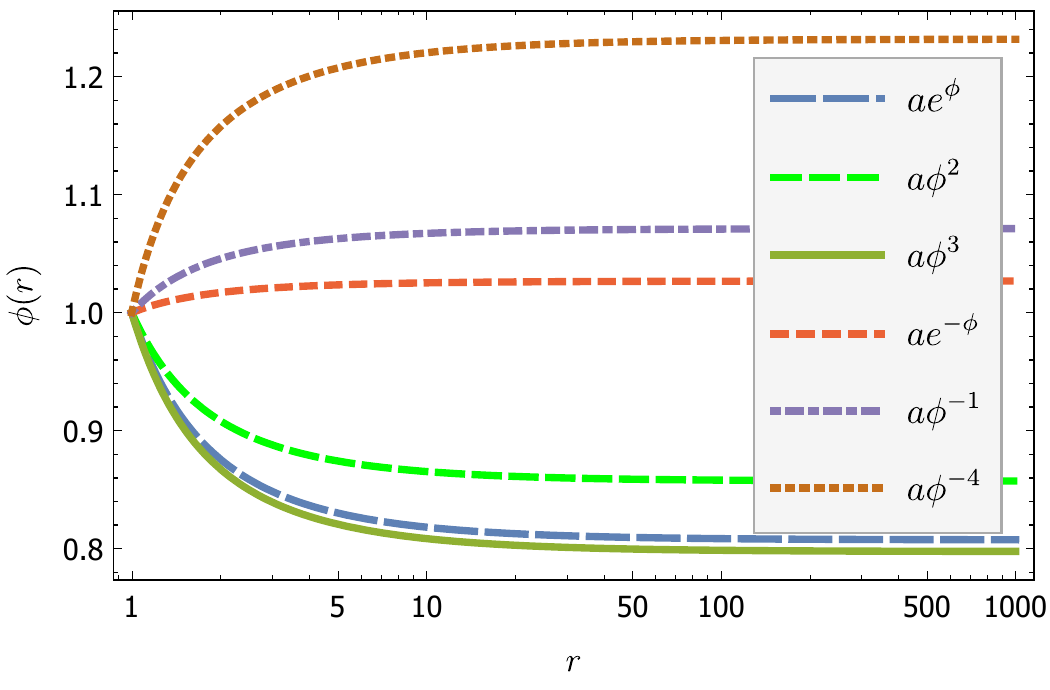}} \hspace*{-0.5cm}
\includegraphics[height=1.75in]{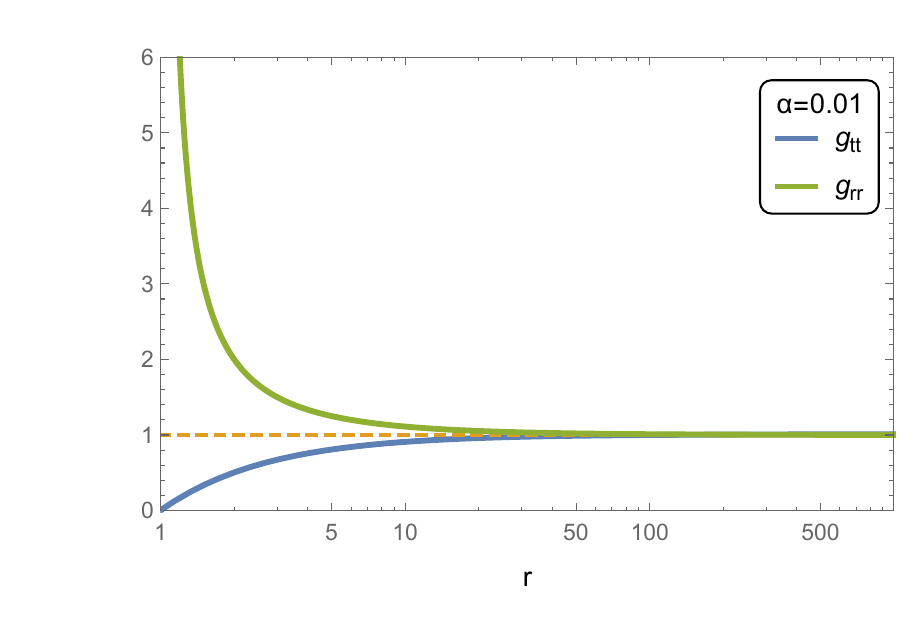} % requires the graphicx package
\caption{The profile of the scalar field for a variety of choices for the coupling function $f(\phi)$ (left plot),
and the two metric functions (in absolute value, right plot) \cite{ABK1}\cite{ABK2}.}
 \label{fig:sol_asym}
\end{figure}

The aforementioned argument was followed also in \cite{DBH1} to prove that the dilatonic theory with
$f(\phi) = \alpha e^{\phi}$ evades Bekenstein's theorem, a result which opened the way to find new black-hole
solutions, the so-called dilatonic black holes. In the context of the present analysis, it is clear that the exact form
of the coupling function is again of no importance. Thus, we set off looking for new black holes in the context of
the theory (\ref{action}) choosing various forms for $f(\phi)$. For each form, we numerically integrated the
system of equations (\ref{fin_sys}), by giving the input values $(\phi_h, \phi'_h)$. The first quantity was 
a free parameter constrained only by the condition (\ref{con_f}) while the second one was uniquely determined
by the regularity constraint (\ref{phi'_con}) of the black-hole horizon. Using this method, every pair of initial
values $(\phi_h, \phi'_h)$ leads to a regular black-hole solution with a non-trivial scalar hair. In this way, we
determined a large number of black-hole solutions with scalar hair for a variety of forms of the coupling
function $f(\phi)$: exponential, odd and even power-law, odd and even inverse-power-law. An indicative
subset of solutions for the scalar field are depicted in the left plot of Fig.~\ref{fig:sol_asym} while, on the right plot, we present
the profile of the two metric functions (in absolute value) with the characteristic form of an asymptotically-flat
black-hole geometry. We should note at this point that our analysis covers both the case of {\it spontaneous
scalarisation} (i.e. the case where the Schwarzschild solution arises as an independent solution) and the case
of {\it natural scalarisation} (where the Schwarzschild solution does not emerge) depending on the particular
choice of the coupling function $f(\phi)$.

%%%%%%%%%%%%%%%%%%%

%%%%%%%%%%%%%%%%%

\begin{figure}[t]
\mbox{\includegraphics[width=2.4in]{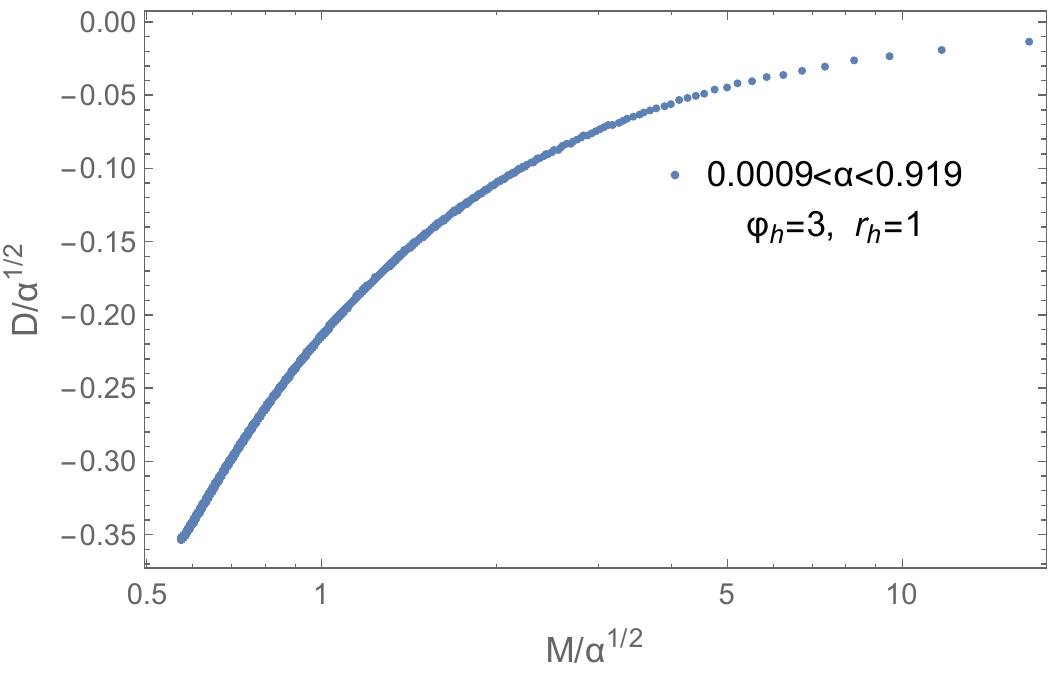}}
\hspace*{-0.0cm}{\includegraphics[width=2.4in]{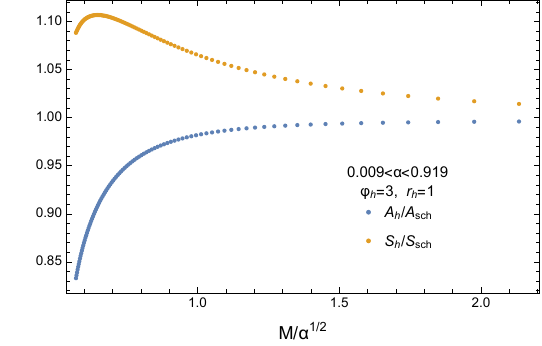}}
\caption{The scalar charge $D$ (left plot), and the ratios $A_h/A_{Sch}$ and $S_h/S_{Sch}$ 
(right plot, lower and upper curve respectively) in terms of the mass  $M$, for
$f(\phi)=a\,/\phi$ \cite{ABK1}.}
\label{fig:prop}
\end{figure}

In Fig.~\ref{fig:prop}, we present some of the properties of the scalarised black holes for the indicative case
of $f(\phi)=a/\phi$. The left plot presents the scalar charge $D$, that characterises the scalar field at infinity. 
As is clear, this quantity is in fact a function of the black-hole mass, a result that renders the scalar hair
secondary. Also, a common characteristic in all cases is that, as the mass of the black hole increases, the
scalar charge decreases and eventually vanishes as our black-hole solution matches the Schwarzschild solution.
In other words, every massive GB scalarised black hole reduces to the Schwarzschild solution. 

%%%%%%%%%%%%%%%%%%%%

The right plot of Fig.~\ref{fig:prop} depicts two important quantities, the horizon area of the black hole, given
by $A_h=4 \pi r_h^2$ and normalised in units of the horizon area of the Schwarzschild solution with the same mass,
and the entropy of the black hole normalised again in units of the entropy of the corresponding Schwarzschild
solution. Starting from the horizon area, we observe that this ratio is always smaller than unity, which means
that all GB black-hole solutions are smaller than their GR analogues. This is due to the effect of the GB term
which exerts a positive (outward) pressure and which can be counterbalanced only by ``squeezing'' further
the available energy/mass distribution thus resulting into smaller black holes. We also observe that the area curve stops
abruptly at its lower end, thus exhibiting the existence of a lower bound on the horizon area and thus on the mass
of the black hole. Beyond this lower value, the black hole ceases to exist --- this feature is due to the bound
(\ref{con_f}) discussed earlier.  The entropy of the black-hole solutions may be computed, for an arbitrary form
of $f(\phi)$, following various methods \cite{KT}\cite{ABK2}, and it is found to be 
%%%%%%%%%%%%%%
\beq
S_h=\frac{A_h}{4} +4 \pi f(\phi_h)\,.
\label{entropy}
\eeq
%%%%%%%%%%%%%%
The profile of the entropy $S_h$ of a GB black hole compared to the one of the Schwarzschild black hole,
$S_{Sch}=A_h/4$, depends strongly on the choice for the particular form of the coupling function. For instance,
for the choice employed in this case, i.e. the inverse linear form, the entropy ratio comes out to be larger than
unity for the entire mass range, a result that renders this particular family of solutions more thermodynamically
stable than the corresponding Schwarzschild black hole. For different forms of the coupling function though,
the curve of the entropy ratio may lie in whole or in parts below unity thus revealing an instability for the entire
or parts of the mass regime. In all cases, as the mass $M$ increases, the entropy ratio always reduces to unity
as the Schwarzschild limit is approached.

The theory (\ref{action}) allows not only asymptotically-flat black-hole solutions but also black holes with
an asymptotically Anti-de Sitter behaviour. This follows if we add to the theory a cosmological constant
bringing the action functional of the theory to the form \cite{BAK}
\beq
S=\frac{1}{16\pi}\int{d^4x \sqrt{-g}\left[R-
\frac{1}{2}\,\partial_{\mu}\phi\partial^{\mu}\phi+f(\phi)R^2_{GB}- 2\Lambda\right]}.
\eeq
%%%%%%%%%%%%%%%%
\smallskip
In this case, the field equations (\ref{set_field}) remain unchanged apart from the
shift 
%%%%%%%%%%%%%%%%%
\beq
T_{\mu\nu} \rightarrow T_{\mu\nu} -\Lambda \,g_{\mu\nu}\,.
\eeq
%%%%%%%%%%%%%%%%
Apart from the change in the spacetime background at large distances, we expect that
the profile of the scalar field will also be affected. Indeed, imposing as before the regularity
of its second derivative $\phi''$ at the black-hole horizon, the field equations lead to the
modified constraint
\beq
\phi'_h=\frac{16 \Lambda r_h \dot f^2\,(\Lambda r_h^2 -3) + \Lambda r_h^5 - r_h^3
\mp \sqrt{C}}{4 \dot f [r_h^2 - \Lambda (r_h^4 -\dot f^2)]}\,, \label{phi'_Ads}
\eeq
where all quantities have been evaluated at $r_h$. The quantity $C$ under the square root
stands for the following combination
\begin{equation}
C=256 \Lambda  \dot f^4_h \left(\Lambda  r_h^2-6\right) +
   32 r_h^2 \dot f^2_h \left(2\Lambda  r_h^2-3\right)+r_h^6 \geq 0\,,
\label{C-def}
\end{equation}
%%%%%%%%%%%%%%
and must always be non-negative for $\phi'_h$ to be real. Under the validity of the constraint
(\ref{phi'_Ads}), the asymptotic form of the solution for the metric function and the scalar field
takes the same functional form as the one given in (\ref{sol_rh}). 

Assuming the presence of a negative cosmological constant, and expecting the spacetime to
assume a form close to that of the Schwarzschild-Anti-de Sitter solution at large distances, we
find that the approximate forms for the metric functions in that regime is
%%%%%%%%%%%%%%%%%%
\bea
e^{A(r)}&=& \left(k-\frac{2M}{r}-\frac{\Lambda_{eff}}{3}\,r^2+\frac{q_2}{r^2}\right)
\left(1+\frac{q_1}{r^2}\right)^2,\label{alfar1}\\[3mm]
e^{-B(r)}&=& k-\frac{2M}{r}-\frac{\Lambda_{eff}}{3}\,r^2 +\frac{q_2}{r^2},\label{bfar1}
\eea
%%%%%%%%%%%%%%%%%%
where $k$, $M$, $\Lambda_{eff}$ and $q_{1,2}$ are arbitrary constants. Regarding the asymptotic
form of the scalar field, this is given by
%%%%%%%%%%%%%%%%
\beq
\phi(r) = \phi_\infty + d_1 \ln r +\frac{d_2}{r^2} +  \frac{d_3}{r^3}+ ...\,, \label{phi-far-Anti}
\eeq
%%%%%%%%%%%%%%
where again $(\phi_\infty, d_1, d_2, d_3)$ are arbitrary constant coefficients. In principle, the scalar field
assumes this form in the case of a linear coupling function, $f(\phi)=a \phi$ \cite{Hartmann}\cite{Brihaye}, however, it
may be shown numerically that it holds in the perturbative limit of small GB coupling constant $a$, for all
forms of the coupling function. We notice that the dominant term in the expression of $\phi$ at large
distances has a logarithmic form and not an $1/r$ dependence; as a result, no scalar charge may be
attributed to any scalarised solution found. In contrast, the coefficient $M$ in the asymptotic form of
the metric functions may be interpreted again as the gravitational mass of the solution.

%%%%%%%%%%%%%%%%%%%
\begin{figure}[t!]
%\hspace*{5.0cm}
\mbox{\includegraphics[height=1.56in]{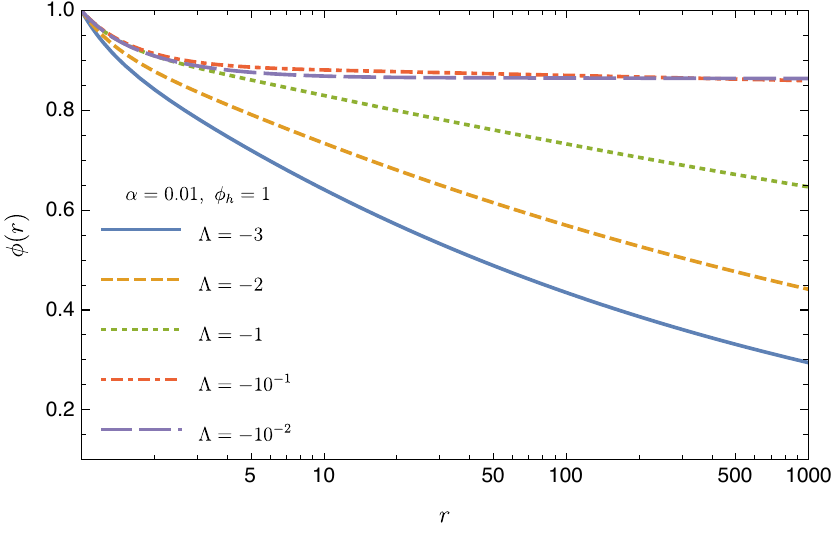}}
{\includegraphics[height=1.58in]{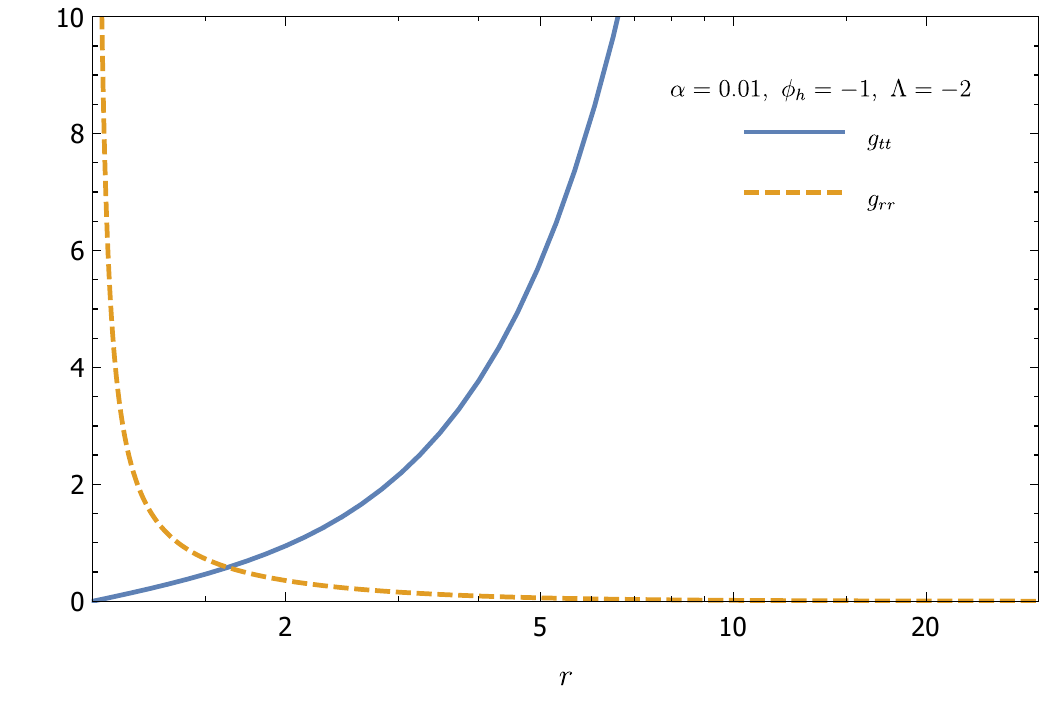}} % requires the graphicx package
   \caption{The solution for the scalar field $\phi$ (left plot), and for the metric components $|g_{tt}|$ and $g_{rr}$
   (right plot) in terms of the radial coordinate   $r$, for $f(\phi)=a \phi^2$ \cite{ABK2}.}
   \label{Fig-met-Lam}
\end{figure}
%
%%%%%%%%%%%%%%%%%

As in the case of zero cosmological constant, a plethora of solutions emerged from the numerical integration
of the set of field equations under the appropriate boundary conditions. Ensuring that the input quantity $\phi'_h$
is given by the expression (\ref{phi'_Ads}), black-hole solutions with an asymptotically Anti-de Sitter behaviour
emerged for all different choices of the coupling function, namely 
$f(\phi)=e^{\pm\phi}, \phi^{\pm 2n}, \phi^{\pm(2n+1)}, \ln\phi,...$ \cite{BAK}. The profiles of both the scalar field
and the metric functions are given in the left and right plot, respectively, of Fig. \ref{Fig-met-Lam}.

We may finally promote the negative cosmological constant to a dynamic potential for
the scalar field \cite{BKP2} as follows
\beq
S=\int d^4 x\,\sqrt{-g}\,\left[\frac{R}{16 \pi G}-
\frac{1}{2}\,\partial_\mu \phi\,\partial^\mu \phi + 
f(\phi)\,R^2_{GB} -2\Lambda\,V(\phi)\right].
\eeq
%%%%%%
We will insist again on the case of the negative coupling constant, i.e. $\Lambda<0$, and consider different choices
for the coupling function $f(\phi)$ and the scalar potential $V(\phi)$. Demanding again the regularity of the black-hole
horizon and repeating the aforementioned analysis, we may determine a similar constraint on the value of $\phi'_h$.
This quantity may be used again an an input parameter, together with $\phi_h$, for the numerical integration of the
field equations, once a specific form for $f(\phi)$ and $V(\phi)$ is chosen. Again, for every possible choice of 
these two scalar functions, black-hole solutions with a regular horizon and a non-trivial $\phi$ emerged. 

\begin{figure}[t]
%\hspace*{5.0cm}
\mbox{\includegraphics[height=1.55in]{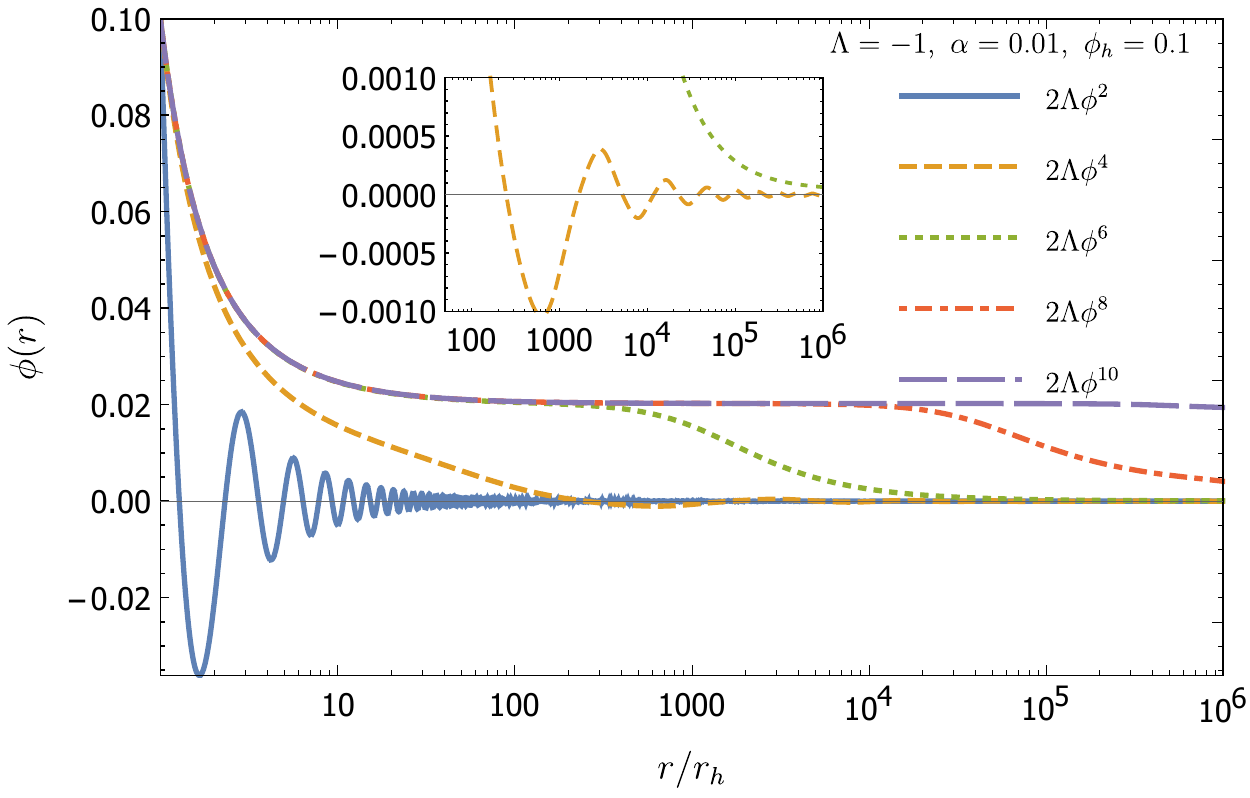}}
\includegraphics[height=1.55in]{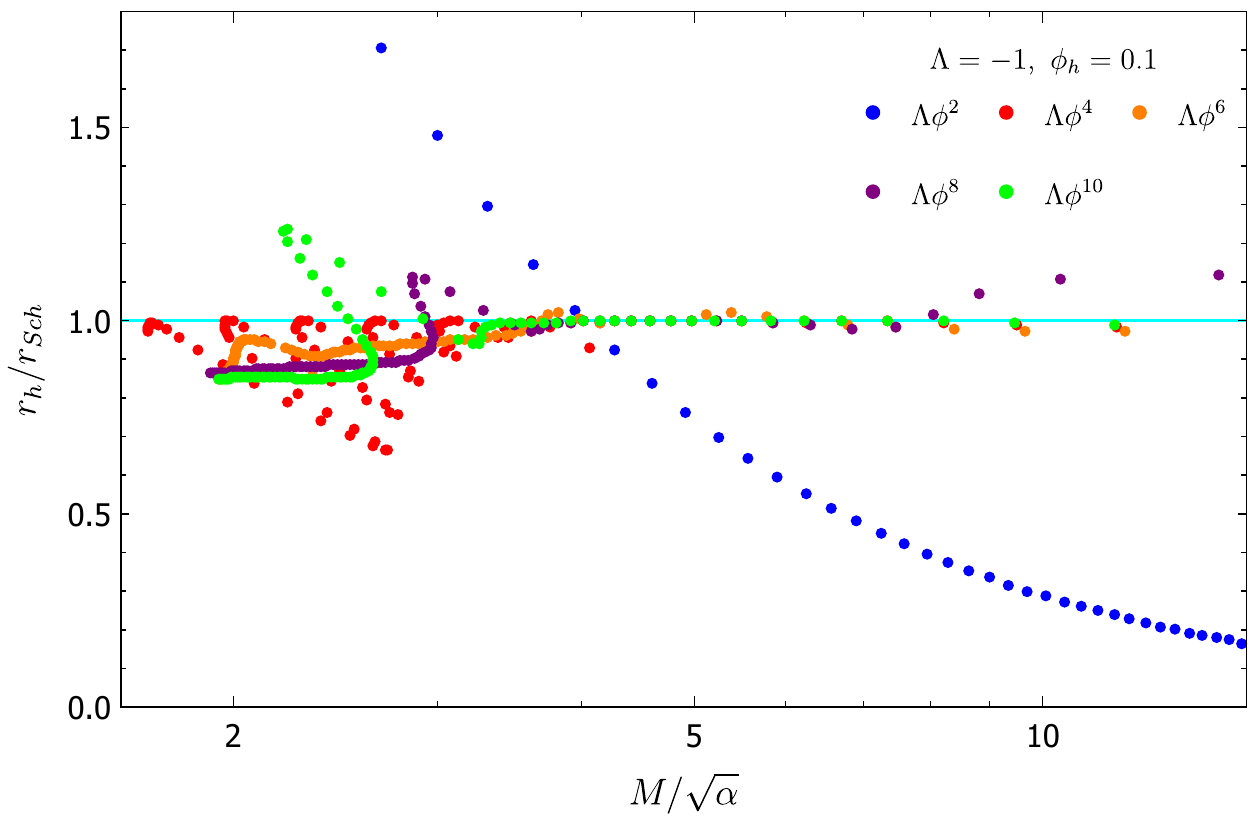} % requires the graphicx package
\caption{The profile of the scalar field in terms of the radial coordinate for 
different (polynomial) choices of the scalar potential (left plot); the lines of existence of the black-hole
solutions for the same choices of $V(\phi)$ (right plot)  \cite{BKP2}.}
\label{fig:dynamic}
\end{figure}

In Fig. \ref{fig:dynamic} (left plot), we present the profile of the scalar field in terms of the radial coordinate for many
different (polynomial) choices of the scalar potential, and for fixed form of the coupling function, i.e. $f(\phi)=a e^\phi$.
We observe that for $\Lambda<0$, the scalar field oscillates around the zero value where it finally relaxes. As the
degree of the polynomial increases, the relaxation time gets longer. In the right plot, we depict the domain of existence
of the black-hole solutions under the same choices. Here, we notice a different pattern for the solutions emerging for
$V(\phi)=\Lambda \phi^2$ from the one assumed for the other polynomial forms. In the latter cases, the black-hole
solutions form branches which are more prominent in the small-mass regime but tend to smooth out and
eventually match the horizon value of the corresponding Schwarzschild solution. In the case of a quadratic scalar
potential, the solutions present a distinctly different pattern: the solutions form a monotonically decreasing line of
existence spanning the whole regime, from large-$r_h$, low-mass black holes to small-$r_h$, large-mass black holes.
The theory therefore includes {\it ultra-sparse} black holes, {\it Schwarzschild-like} black holes and
{\it ultra-compact} black holes. The study of the entropy reveals a similar behaviour in terms of the mass as
the horizon radius; as a result, the ultra-sparse black holes are found to be more thermodynamically stable
compared to the Schwarzschild-like black holes, and these in turn to be  more thermodynamically stable
compared to the ultra-compact black holes.

%\smallskip
%%%%%%%%%%%%%%%%%%%

%%%%%%%%%%%%%%%%%

The discussion above summarizes only a small part of the black-hole solutions that have been found
in the context of gravitational theories containing the GB or related terms. There is in fact a huge 
literature in this topic and the interested reader may find more results in \cite{Doneva}-\cite{BHR}.

%%%%%%%%%%%%  SLIDE 3 %%%%%%%%%%%%%%%%%%%%
\subsection{Wormholes in Einstein-Scalar-GB Theory}

As we stated in Section 2, General Relativity does not accommodate {\it traversable} wormholes. We also
claimed that the Ellis-Bronnikov solution, which emerges in the context of the Einstein-scalar theory, is supported
in fact by a ghost scalar field. Is there a particular feature of the wormhole geometry that makes it incompatible
with GR or simple extensions of it? Actually, there is. 

To demonstrate this, we will employ the Morris-Thorne method for the construction of wormhole solutions
\cite{MT}. This method also incorporates the alternative approach mentioned in Section 2, in which traversable
wormhole solutions may emerge only in the absence of a horizon or singularity. According to the analysis
of \cite{MT}, a traversable wormhole may be described by the following line-element 
\beq
ds^2=-e^{2 \Phi (r)}\,dt^2 + \left(1-\frac{b(r)}{r}\right)^{-1} dr^2 +
r^2\,(d\theta^2 +\sin^2\theta\,d\varphi^2)\,. \label{MT}
\eeq
%%%%%%%%
The {\it red-shift} function $\Phi$ must be everywhere finite. The {\it shape} function $b$ must satisfy
$1-b/r\geq 0$ throughout spacetime, and is allowed to vanish at a single point, i.e. at $r=b=b_0$, where the
throat of the wormhole is located. At infinity, $\Phi$ and $b/r$ must both vanish to recover the asymptotically flat regime. 

\begin{figure}[t!]
\hspace*{2.5cm}
\includegraphics[height=2.0in]{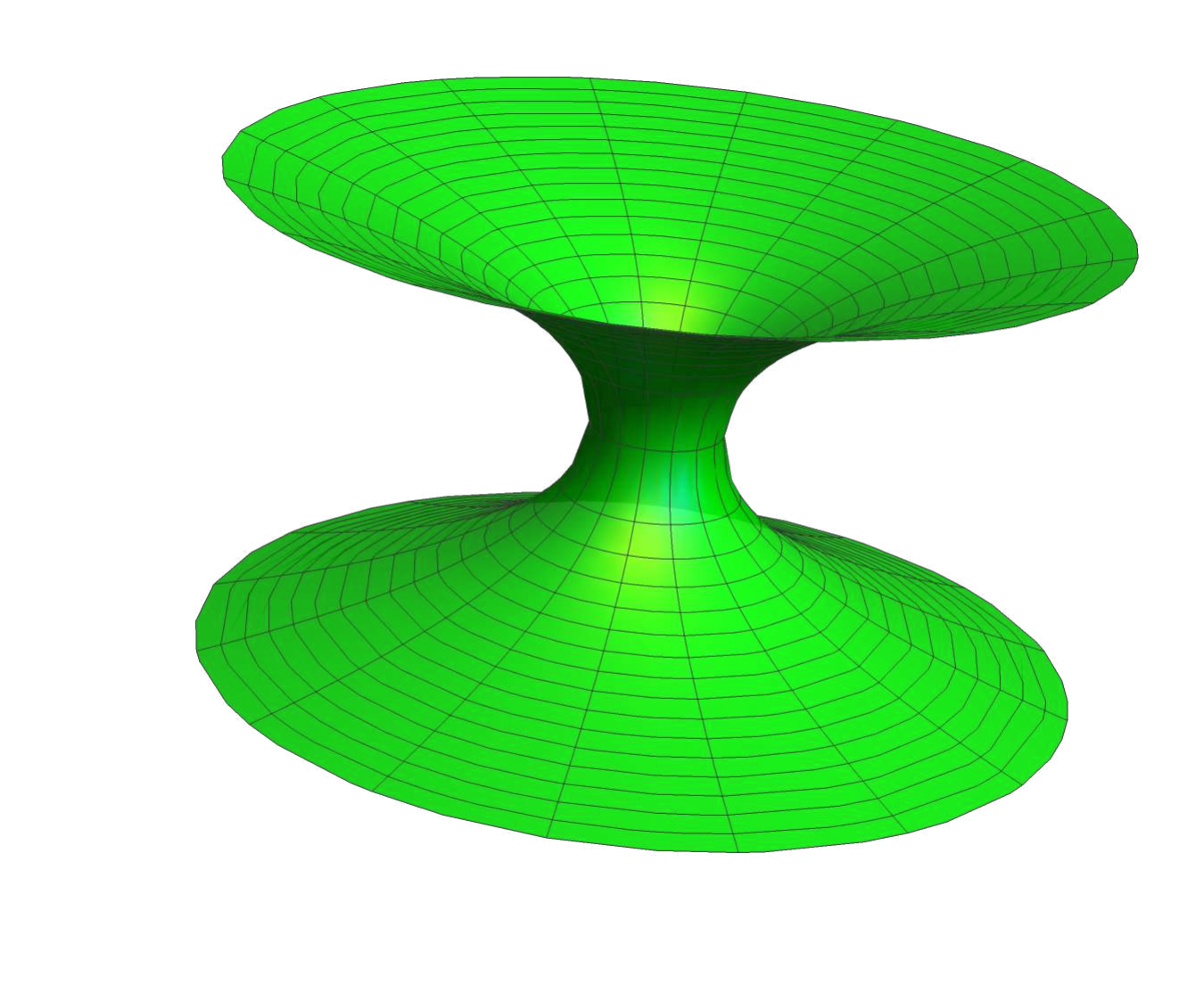}
\caption{The embedding diagram of a typical traversable wormhole.}
\label{fig:embed}
\end{figure}

To study the geometric structure of the above solution, we construct its embedding diagram. To this end,
we equate the line-element of the 2D spacelike equatorial surface, obtained for $t=const.$ and $\theta=\pi/2$,
with the one of a 3D Euclidean space:
%%%%
\begin{equation}
    \frac{dr^2}{1-b/r}+ r^2\,d \varphi^2 \equiv dz^2 + d\rho^2 + \rho^2\,d\varphi^2=
    \left[ \left(\frac{dz}{d\rho}\right)^2 +1 \right] d\rho^2 +\rho^2\,d\varphi^2\,.
    \label{embedMT}
\end{equation}
%%%%%%
From the above, we obtain immediately that $\rho=r$. We demand that both line-elements describe the 
same geometry and equate the coefficients of $dr^2$ in \eqref{embedMT}. Then, the ``fictitious" $z$ coordinate
is given by the relation
%%%
\begin{equation}
    \frac{dz}{dr}= \pm \left(\frac{r}{b}-1\right)^{-1/2} \,\Rightarrow \, z(r)=\pm \int_{b_0}^r \left(\frac{r}{b}-1\right)^{-1/2} dr\,,
\end{equation}
and describes the lift of the ($r,\varphi$)-plane due to the curvature of spacetime. At infinity, as $b/r \rightarrow 0$,
we obtain $dz/dr \rightarrow 0$, and the embedding surface becomes parallel to the ($r,\varphi$)-plane exhibiting
no lift due to the flatness of this regime. In contrast, near the throat, the slope diverges and this signifies that the
embedding surface at this point is vertical. Therefore, $\{\rho(r), z(r)\}$ is a parametric representation of a slice of
the embedded $\theta = \pi/2$-plane for a fixed value of the $\varphi$ coordinate, while the corresponding surface
of revolution is the 3-D representation of the wormhole's geometry. The typical embedding diagram of a wormhole
is given in Fig. \ref{fig:embed}.

%%%%%%%%%%%%%%%%%%%
%\begin{center}
%\end{center}
%
%%%%%%%%%%%%%%%%%

One could invert the function $z(r)$ to obtain $r(z)$. Then, from Fig. \ref{fig:embed}, it is clear that  $r(z)$ possesses a minimum
at the throat and then ``flares outwards'' as we approach the asymptotically flat region. Therefore, the existence of a throat
is mathematically ensured if the following two conditions hold
%%%%%%%
\begin{equation}
\frac{dr}{dz}\biggr|_{r_0}=0\,, \qquad \quad \frac{d^2r}{dz^2}\biggr|_{r_0} >0\,.
\end{equation}
%%%%%%%%
The first condition is satisfied since, as we showed above, $dz/dr \rightarrow 0$ at $r=r_0$. The second condition
takes the form \begin{equation}
  \frac{d^2r}{dz^2} = \frac{b- rb'}{2b^2}>0\,, \label{flaring-out}
\end{equation}
and is known as the {\it flaring-out} condition of the wormhole. Therefore, in order to have a throat, the shape
function $b(r)$ should satisfy $b- rb'>0$, which fully justifies its name. 

In fact, the flaring-out condition leads to the violation of energy conditions in a wormhole background since it imposes
a certain behaviour on the shape function and, through Einstein's equations, on the matter content of the theory.
Let us focus on the Null Energy Condition (NEC) which has the form: 
\begin{align}
T_{\mu\nu}n^\mu n^\nu =- g_{tt}\,(T^r_r - T^t_t) = - g_{tt}\,(\rho+p_r) \ge 0,
\end{align}
where $T_{\mu\nu}$ is the energy-momentum tensor and $n^\mu$ any null vector satisfying $n^\mu n_\mu=0$.
For $g_{tt}<0$, the above demands that $\rho+p_r \ge 0$. However, through Einstein's equations and using the
line-element (\ref{MT}), we obtain
\beq
8 \pi G \, (\rho+p_r) = G^r_r - G^t_t = - \frac{(b - r b')}{r^3}\Bigl |_{r=r_0} <0\,,
\eeq
where we used the flaring-out condition (\ref{flaring-out}). As a result, the wormhole geometry may be supported
by some distribution of matter which violates the energy conditions and instead satisfies the condition $\rho+p_r <0$.
That is why pure GR cannot support a traversable wormhole nor can the Einstein-scalar theory with a physical
scalar field do it either. 

It becomes therefore clear that a more involved gravitational theory beyond GR is necessary in order to
support viable wormhole solutions. Can we find such a physical theory? The Einstein-scalar-GB theory is
a theory which violates the energy conditions, as the evasion of no-hair theorems clearly demonstrated. 
However, this violation is not caused by the introduction of some form of exotic matter but by the direct
coupling of a physical scalar field to the quadratic, gravitational GB term. Can this violation then support
wormhole solutions?

In \cite{DBH1}, where the dilatonic black holes were first discovered, in the context of the EsGB theory
with an exponential coupling function, $f(\phi)=a\,e^\phi$, another class of solutions were
presented with the line-element having the form of a Morris-Thorne solution. By applying a coordinate
transformation \cite{KKK1}\cite{KKK2}, this was written as 
\beq
ds^2=-e^{A(\ell)} dt^2 + e^{B(\ell)} d\ell^2 +(\ell^2 +r_0^2)\,(d\theta^2 +\sin^2\theta d\varphi^2)\,,
\eeq
where $\ell$ is a new spacelike coordinate ranging in the interval $(-\infty, \infty)$. The above line-element
describes a wormhole connecting two asymptotically-flat regimes and with a radius throat $r_0$ at $\ell=0$.
In this new coordinate system, both metric functions and the scalar field assume regular forms in the regime
close to the throat, namely
\beq
e^{A(\ell)}=a_0+a_1 \ell +...\,, \quad e^{B(\ell)}=b_0 + b_1 \ell +...,, \quad \phi(\ell) = \phi_0 + \phi_1 \ell +...
\eeq
%%%%%%%
On the other hand, at large distances, we obtain power-law expansions in terms of $(1/r)$ 
\beq
e^{A} \simeq 1-\frac{2M}{\ell} + ..., \qquad e^{B} = 1+\frac{2M}{\ell} + ...\,, \qquad
\phi \simeq \phi_\infty + \frac{D}{\ell} + ...\,,
\eeq
where $M$ and $D$ are the mass and scalar charge of the wormhole. Note that these expansions are
identical to the ones for a black-hole solution with the exception that here $M$ and $D$ are independent
quantities. 

%%%%%%%%%%%%%%%%%%%
\begin{figure}[t!]
%\hspace*{5.0cm}
\begin{center}
\mbox{\includegraphics[height=3.5in]{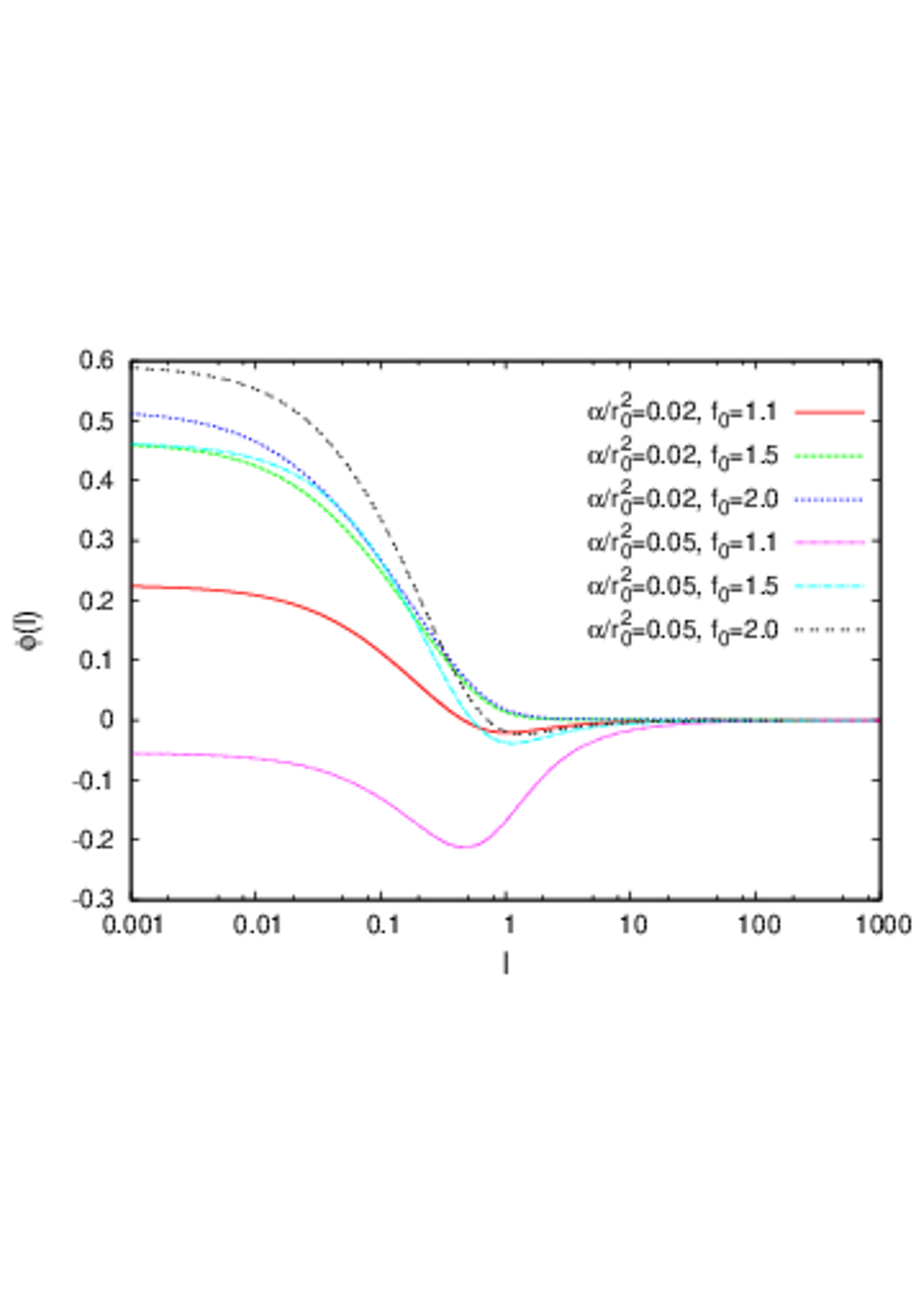}} 
{ \vskip -2.0cm \includegraphics[height=2.0in]{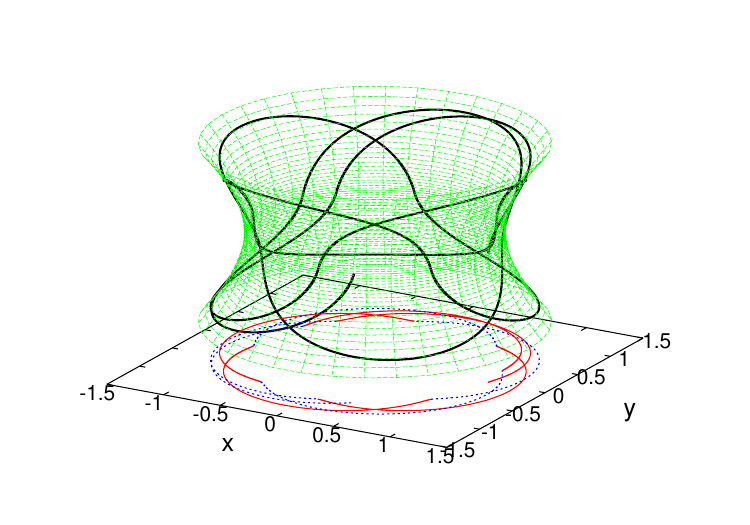}} % requires the graphicx package
\caption{The solutions for the scalar field for a family of dilatonic wormholes (upper plot), and particles trajectories in
the background of a dilatonic wormhole (lower plot) \cite{KKK1}\cite{KKK2}.}
 \label{fig:worm_num}
\end{center}
\end{figure}
%
%%%%%%%%%%%%%%%%%

Employing these asymptotic solutions as boundary conditions, the set of field equations (\ref{set_field}) were
numerically solved again, this time in quest of wormhole solutions \cite{KKK1}\cite{KKK2}. And such solutions were
indeed found, with the spacetime background having the geometry depicted in Fig. \ref{fig:embed} and the
scalar field assuming the profile presented on the upper plot of Fig. \ref{fig:worm_num}. However, a true singularity
was lurking behind the throat, at the negative $\ell$ regime. Thus, a traversable, symmetric solution was built by
cutting the spacetime at $\ell=0$ and gluing the positive regular $\ell$-regime with its mirror image along $\ell <0$.
At the gluing point $\ell=0$, cusps were created whose presence could be justified only by introducing  a 
distribution of a (non-exotic!) perfect fluid and a gravitational source term around the throat. This distribution
is described by the action functional \cite{KKK1}\cite{KKK2}
\beq
S_3= \int d^3 x \sqrt{h} \,(\lambda_1+ \lambda_0 e^{\phi} \tilde R)\,,
\eeq
where $(\lambda_1, \lambda_2)$ are constants and $\tilde R$ is the scalar curvature of the 3-dimensional induced
spacetime at $\ell=0$. Under the above construction, a regular wormhole solution is constructed which can be 
traversed both by null and timelike particles. A set of particle trajectories, which start from one asymptotically-flat
regime, traverse the throat, reach the asymptotically-flat regime on the other side of the wormhole and then 
travel back to the starting point, are presented on the lower plot of Fig. \ref{fig:worm_num}.

%%%%%%%%%%%%%%%%%%
\begin{figure}[t!]
\hspace*{-1.0cm}
\mbox{\includegraphics[height=1.8in]{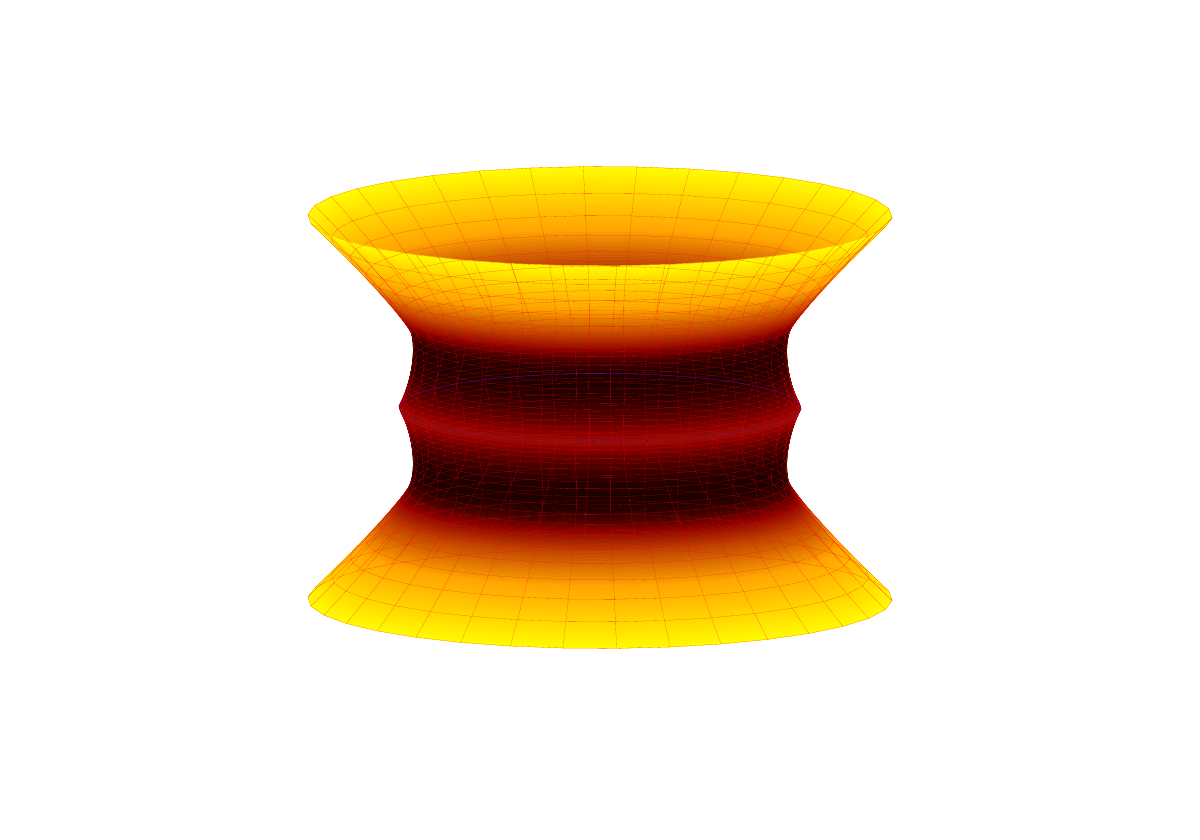}} %\hspace*{-0.5cm}
{\includegraphics[height=1.8in]{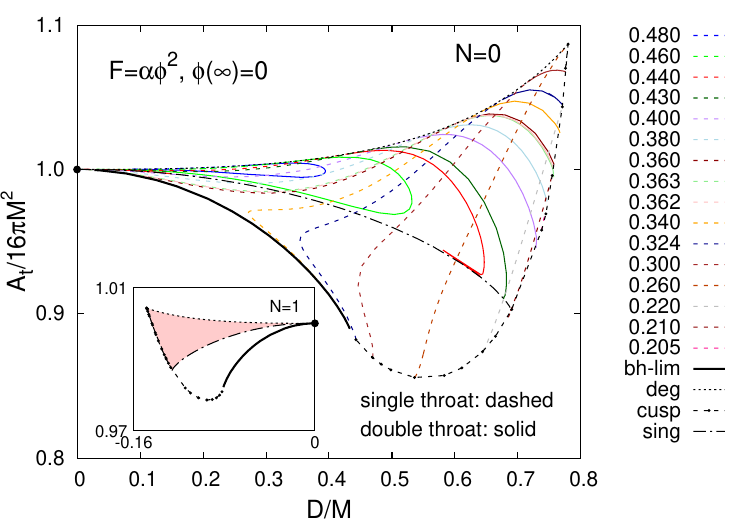}} % requires the graphicx package
\caption{A wormhole solution with an equator and two throats (left plot), and the domain of existence
of wormhole solutions in the case where $f(\phi)=a \phi^2$ (right plot) \cite{ABKKK}.}
  \label{fig:embed-double}
\end{figure}
%
%%%%%%%%%%%%%%%%%

All the above results were derived in the context of the dilatonic Einstein-scalar-GB theory, that is with an
exponential coupling function. Can we generalise this analysis and produce similar wormhole solutions in
the context of the EsGB theory with alternative forms for $f(\phi)$, as in the case of black holes? In the
context of the general theory (\ref{action}), we looked indeed for viable wormhole solutions following a similar
approach as above. Apart from the form of the coupling function, we incorporated also a different form of
line-element, namely the following \cite{ABKKK}
%%%%%%%%
\beq
ds^2=-e^{A(\eta)} dt^2 + e^{\Gamma(\eta)} \left[d\eta^2 
+(\eta^2 +\eta_0^2)\,(d\theta^2 +\sin^2\theta d\varphi^2)\right].
\eeq
%%%%%%%%
This new ansatz modifies the conditions for the existence of a sole minimum in the function $r(z)$ in the
embedding diagram, and allows for multiple extremal points and thus for a much richer topology: local maxima
correspond to the so-called {\it equators} whereas local minima give rise to {\it multiple throats}. 

The metric functions and scalar field assume regular forms both near the throat and the far asymptotic
regime. The numerical integration yielded a large number of wormhole solutions, for every form of $f(\phi)$
used, with an asymptotically-flat behaviour and with a single throat or with a double throat and an equator. 
The embedding diagram of a wormhole in the latter case is shown on the left plot of Fig. \ref{fig:embed-double}.
All solutions were made traversable by employing the same cut \& paste technique, where the spacetime 
with $\ell>0$ was  glued at $\ell=0$ with its mirror image along $\ell <0$. As before,  a distribution of regular
matter around the throat suffices to justify the cusps at the gluing point. On the right plot of 
Fig. \ref{fig:embed-double}, we present the domain of existence of wormhole solutions in the indicative
case of quadratic coupling function, $f(\phi)=a \phi^2$.  All EsGB wormhole solutions are bounded by
the corresponding black-hole solutions which form the left boundary of the domain.

For additional works on wormhole solutions arising in the context of generalised gravitational theories
involving scalar fields or higher-curvature terms see, for instance, \cite{Bronnikov:1997}-\cite{Ibadov:2021}.

%%%%%%%%%%%%  SLIDE 16 %%%%%%%%%%%%%%%%%%%%
\subsection{ Particle-like Solutions in Einstein-Scalar-GB Theory}

Let us finally address the last class of compact objects, the gravitational particle-like solutions. As was mentioned,
in the context of the minimal Einstein-scalar theory, the corresponding Fisher/Janis-Newman-Winicour-Wyman
particle-like solution \cite{Fisher} is irregular. Can we hope to find particle-like solutions which however behave
better in the context of the EsGB theory?

With this in mind, we looked for static, spherically-symmetric solutions which describe a regular spacetime, with
no singularities or horizons and no throats as well \cite{KKK_part1}\cite{KKK_part2}. We chose to work with the following
form of line-element 
\beq
ds^2=-e^{A(r)} dt^2 + e^{\Gamma(r)}\left[dr^2 +r^2\,(d\theta^2 +\sin^2\theta d\varphi^2)\right],
\eeq
with two unknown metric functions, $A(r)$ and $\Gamma (r)$, as usual. The above form was inspired by the
one employed in our quest for wormhole solutions and resulted, as we saw, in a much richer topology of these
solutions around the throat. Here, we will use a similar line-element in the hope that solutions with features
not allowed or often observed, when more restrictive ansatzes are used, will now emerge. 

The set of field equations of the theory (\ref{set_field}) must now be solved for the three unknown functions, 
the two metric functions and the scalar field $\phi(r)$. If we substitute the above ansatz for the line-element
in the field equations, 
we obtain four, coupled, ordinary differential equations which can further be reduced to a system of only two
independent equations for $A(r)$ and $\phi(r)$. The remaining metric function $\Gamma(r)$ can be
determined once the solutions for $A(r)$ and $\phi(r)$ are found. 

\begin{figure}[t!]
%\hspace*{1.0cm}
\mbox{\includegraphics[height=1.7in]{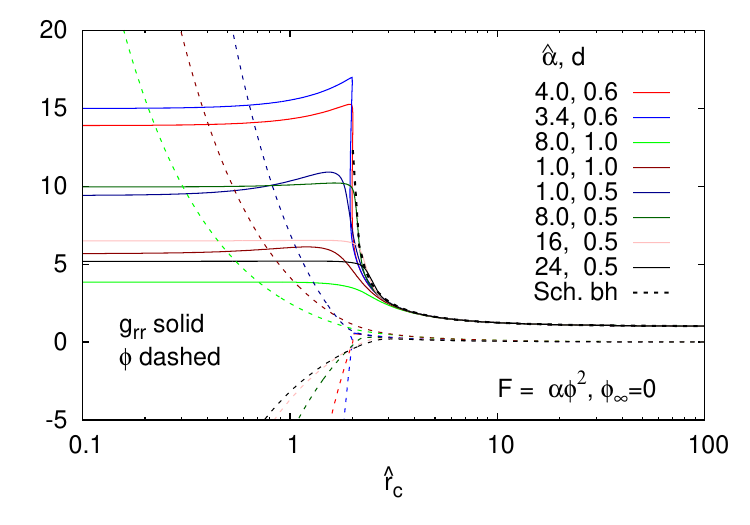}} %\hspace*{-0.5cm}
{\includegraphics[height=1.7in]{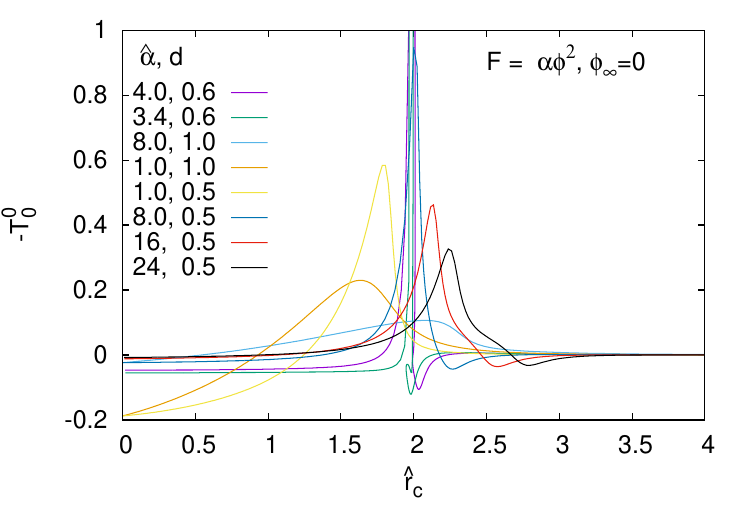}} % requires the graphicx package
\caption{The solutions for the metric and scalar field (left plot) and the corresponding energy-density
(right plot) in the case of the quadratic coupling function  \cite{KKK_part1}\cite{KKK_part2}.}
\label{fig:particles}
\end{figure}

The set of the two independent, second-order differential equations was solved numerically under the proper
boundary conditions. We demanded a regular form for both functions over the entire radial regime, an
asymptotically-flat behaviour at large distances and a smooth solution close to the origin. At large distances,
we obtained indeed an asymptotically-flat behaviour, which was identical to the one for black holes and
wormholes, namely
%%%%%%%
\beq
e^{A(r)} \simeq 1-\frac{2M}{r} + ..., \qquad \phi \simeq \phi_\infty - \frac{D}{\ell} + ...\,,
\eeq
where $M$ and $D$ are once again the mass and the scalar charge of the solution. The expansion near
the origin of the radial coordinate was found to be more involved and to depend on the form of the
coupling function $f(\phi)$. For instance, for a quadratic form, $f(\phi)=\alpha \phi^2$, we find the
asymptotic solution
\beq
A(r)=A_0 + A_2 r^2 + A_3 r^3 + \dots\,,
\eeq
\beq
\phi=-\frac{c_0}{r} +\phi_0 +\phi_1 r + \phi_2 r^2 + \phi_3 r^3 + \dots\,.
\eeq
We readily observe that the metric is indeed regular near the origin but the scalar-field expansion
features a singular term with a divergent behaviour at small $r$. In fact, the presence of this term seems
to be a generic feature of the expansion of the scalar field near the origin independently of the form of
the coupling function. The numerical integration of the system of field equations led to the complete
solution which, as expected, interpolated between the two asymptotic forms. The profile of the complete
solutions for the metric and scalar field are depicted, for the quadratic coupling function, on the left plot
of Fig. \ref{fig:particles} \cite{KKK_part1}\cite{KKK_part2}. 

%%%%%%%%%%%%%%%%%%%
%
%%%%%%%%%%%%%%%%%

In order to investigate the consequences of the singular term, present in the expansion of the scalar
field near the origin, we computed all gravitational scalar invariants as well as the components of the
energy-momentum tensor $T_{\mu\nu}$. Despite the singularity in $\phi$, all these quantities were
found to be finite over the entire radial regime. The exact values of the energy-density and pressure 
at the origin are given below:
%%%%%
\beq
\rho(0)=-\frac{3}{32 \alpha}\,, \qquad p(0)=\frac{2}{32 \alpha}\,.
\eeq
%%%%%%
We may therefore conclude that the singularity of the scalar field at the origin has no physical
consequence, and that the situation resembles more the behaviour of the Coulomb potential which also
diverges at the origin of the coordinate system. 

%%%%%%%%%%%%%%%%%%%
\begin{figure}[t!]
%\hspace*{1.0cm}
\mbox{\includegraphics[height=1.7in]{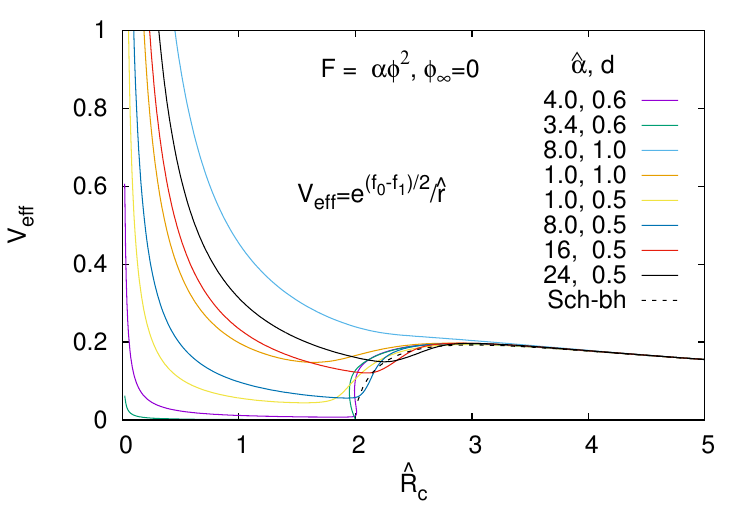}} %\hspace*{-0.5cm}
{\includegraphics[height=1.7in]{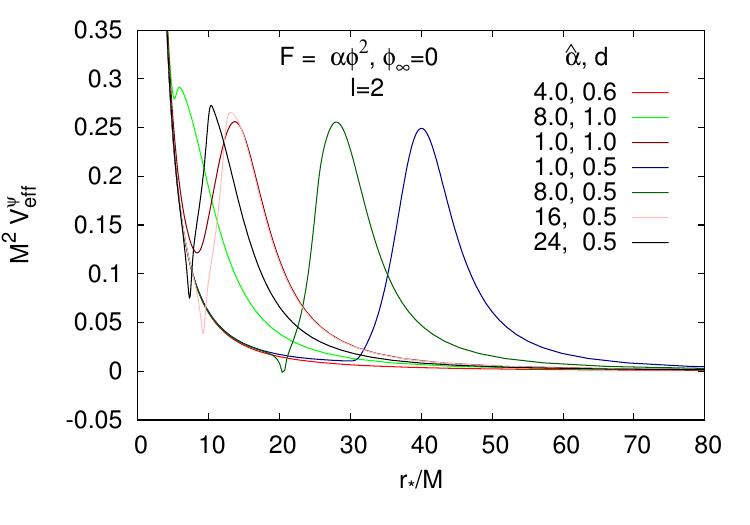}} % requires the graphicx package
\caption{The effective potential for null particles (left plot) and for test scalar particles (right plot), for a 
quadratic coupling function \cite{KKK_part1}\cite{KKK_part2}.}
 \label{fig:light}
\end{figure}
%
%%%%%%%%%%%%%%%%%

The  complete profile of the energy-density $\rho$ of an indicative particle-like solution is depicted on the
right plot of Fig. \ref{fig:particles}. We observe that it has a regular, shell-like behaviour, and vanishes quickly 
at a very small distance from the origin. We may therefore conclude that our gravitational particle-like solutions
are more {\it bubble-shaped} than point-like, and clearly qualify as {\it ultra-compact} objects (see also
\cite{Cardoso:2014}-\cite{Brihaye:2019}).

The propagation of test particles in the background of the aforementioned gravitational bubble-shaped solutions 
may lead to interesting observable phenomena. The geodesics for both null and timelike particles may be found
from their Lagrangian ${\cal L}$  given by the expression
\begin{equation}
2 {\cal L}=g_{\mu\nu} \dot x^\mu \dot x^\nu 
=  -e^{f_0} \dot t^2 +e^{f_1}\left[\dot r^2 
+r^2\left( \dot \theta^2+\sin^2\theta \dot \varphi^2\right) \right]  = - \epsilon \ ,
\label{lag}
\end{equation}
where $\epsilon=0$ and 1 for massless and massive particles, respectively. Employing the two conserved
quantities, namely the energy $E=-e^{f_0} \dot t$ and the angular-momentum $L=e^{f_1} r^2 \dot \varphi$
of the particle, and considering motion in the equatorial plane ($\theta=\pi/2$), we obtain the following
equation for null particles 
\begin{equation}
e^{f_0+f_1} \dot r^2=(E+L V_{\rm eff})(E-L V_{\rm eff}),
\label{Veff_photons}
\end{equation}
where $V_{\rm eff}=e^{(f_0-f_1)/2}/r $ is the effective potential for photons. We depict its form in terms
of the scaled circumferential coordinate $\hat{R}_c=e^{\Gamma/2} r$ in Fig.~\ref{fig:light},
for the coupling function $F=\alpha\phi^2$. Whereas, for the less compact particle-like solutions,
the effective potential is monotonic, for the more compact solutions, the effective potential $V_{\rm eff}$
is characterised by a pair of extrema. These extrema correspond to circular orbits and thus lead to
the presence of {\it  light rings} around these ultra-compact solutions. Bound orbits emerge also for
massive particles with radii smaller than or comparable to the radius of the bubble. In fact, it is found
that particles may pass over the origin ($r=0$) or even rest there without encountering any divergence 
from the singular term in the expression of the scalar field. 

We also considered the propagation of a scalar test particle $\Psi$ in this background. Starting from
the equation $\Box \Psi=0$ and expanding the scalar field in spherical harmonics $Y^m_l(\theta, \varphi)$, 
\begin{equation}
\Psi(t,r,\theta,\varphi)=\sum_{l,m} \psi_{l,m}(t,r)\,e^{-f_1/2} Y^m_l(\theta, \varphi)/r \ ,
\label{Psi}
\end{equation}
we obtain the reduced equation 
\begin{equation}
(\partial_t^2- \partial^2_{r_*} +V^{\psi}_{\rm eff})\,\psi_{l,m}(t,r)=0 \,,
\end{equation}
where $r_*$ is the tortoise coordinate defined through the relation $dr_*= e^{(f_1-f_0)/2} dr$,
and $V^{\psi}_{\rm eff}$ is the scalar-field effective potential
\begin{equation}
V_{\rm eff}^{\psi}=e^{f_0-f_1} \left[\frac{l(l+1)}{r^2}+ \frac{2 (f_1'+f_0') +
r f_1'f_0'+2r f_1 ''}{4r}\right] \ .
\end{equation}
The form of the scaled effective potential $M^2 V^\psi_{\rm eff}$ versus the scaled tortoise coordinate $r_*/M$
is presented on the right plot of Fig.~\ref{fig:light}, again for a quadratic coupling function. We observe that,
for scalar waves with $l>0$, the effective potential features an infinite barrier that resides at the origin and
a finite local barrier at a larger value of $r_*$. An incoming scalar-wave with $l>0$ will be partially transmitted
through the finite local barrier and will then undergo a perpetual process of full and partial reflection from the
infinite and finite barrier, respectively, thus producing a series of {\it echoes} in the wave signal.

%%%%%%%%%%%%  SLIDE 16 %%%%%%%%%%%%%%%%%%%%
\section{Compact Objects  in (beyond) Horndeski Theory}
%%%%%%%%%%%%%%%%%%%%%%%%%%%%%%%%%%%%%%%

The Einstein-scalar-GB theory is a special case of a more general tensor-scalar theory of a single scalar
degree of freedom with field equations containing only up to 2nd-order derivatives, the {\it Horndeski theory} 
\cite{Horndeski}. The latter is defined through the Lagrangian
$$S_{\rm H} = \displaystyle\int {d}^4x \sqrt{-g} \,\left(\mathcal{L}_2+\mathcal{L}_3+\mathcal{L}_4+\mathcal{L}_5\right)$$

\vskip -0.4cm
with
\begin{align}
\mathcal{L}_2 &=G_2(X)\,, \qquad  \mc{L}_3 =-G_3(X) \,\Box \phi  \nonumber,
\\[1mm]
\mc{L}_4 &= G_4(X) R + G_{4X} \left[ (\Box \phi)^2 -\nabla_\mu\pd_\nu\phi \,\nabla^\mu\pd^\nu\phi\right] ,\nonumber
\\
\begin{split}
\mc{L}_5 &= G_5(X) G_{\mu\nu}\nabla^\mu \pd^\nu \phi - \frac{1}{6}\, G_{5X} \big[ (\Box \phi)^3 - 3\,\Box \phi\, \nabla_\mu\pd_\nu\phi\,\nabla^\mu\pd^\nu\phi
\\
&\quad + 2\,\nabla_\mu\pd_\nu\phi\, \nabla^\nu\pd^\rho\phi\, \nabla_\rho\pd^\mu\phi \big]. \nonumber
\end{split}
\end{align}
The coupling functions $G_i$ depend on the scalar field $\phi$ only through its kinetic term
$X \equiv -\nabla_\mu \phi \nabla^\mu \phi/2$, thus the above theory is the special, {\it shift-symmetric}
Horndeski theory. The symbol $G_{iX}$ above denotes the derivative of the function $G_i$ with respect to $X$.

Although the theory is more complicated compared to the Einstein-scalar-Gauss-Bonnet theory and the
determination of analytic solutions therefore seems unlikely, in fact the presence of all the additional terms
and the freedom we have in choosing the exact form of the coupling functions $G_i$ of the theory can help
in deriving new analytic solutions. Indeed, whereas this possibility does not exist in the context of the EsGB
theory, a wise choice for the form of $G_i$ in Horndeski theory can turn the set of field equations into an
integrable set of equations, which can then be solved analytically.  For instance, choosing the following coupling functions
\beq
G_2 =8 \alpha X^2\,, \quad G_3=-8 \alpha X\,, \quad G_4=1+ 4\alpha X\,, \quad
G_5=-4 \alpha \ln|X|\,,
\eeq
%%%%%%%
a static, spherically symmetric black-hole solution of the form
\beq
ds^2=-h(r)\,dt^2 + \frac{dr^2}{f(r)} +r^2 d\Omega^2\,, \label{metric_non_homo}
\eeq
with $h(r)=f(r)$, was analytically derived in the context of the Horndeski theory with the metric function and
scalar field having the following explicit forms \cite{Lu_Pang}\cite{Hennigar}
%%%%%%%
\beq
h(r)=1+\frac{r^2}{2 \alpha}\left(1-\sqrt{1+\frac{8\alpha M}{r^3}}\right), \qquad 
\phi'=\frac{\sqrt{h}- 1}{r \sqrt{h}}\,, \label{Lu_Pang}
\eeq
respectively. The above solution describes an asymptotically-flat black hole with a non-trivial scalar field and a horizon
at $r_H=M+\sqrt{M^2-\alpha}$, where $M$ is the black-hole mass and $\alpha$ the coupling constant of the theory.

The Horndeski theory is certainly not the end of the road when it comes to generalised tensor-scalar theories. 
One could extend this theory to {\it beyond Horndeski} by adding the following well-known terms 
\beq
\mathcal{L}^{\rm bH}_4=F_4(X)\varepsilon^{\mu\nu\rho\sigma}\,\varepsilon^{\alpha\beta\gamma}_{\,\,\,\,\,\,\,\,\,\,\,\sigma}\,\partial_\mu\phi\,
\partial_\alpha\phi\,\nabla_\nu\partial_\beta\phi\,\nabla_\rho\partial_\gamma\phi\,
\eeq
\beq
\mathcal{L}^{\rm bH}_5=F_5(X)\varepsilon^{\mu\nu\rho\sigma}\,\varepsilon^{\alpha\beta\gamma\delta}\,\partial_\mu\phi\,\partial_\alpha\phi\,\nabla_\nu\partial_\beta\phi\,\nabla_\rho\partial_\gamma\phi\,\nabla_\sigma\partial_\delta\phi\,.
\eeq
Despite the presence of the above terms, the theory remains free of ghosts provided that the coupling functions
$(G_4, G_5)$ and $(F_4, F_5)$ of the extended theory satisfy the constraint \cite{BenAchour:2016}
%%%%%%
\beq
X G_{5X} F_4 = 3 F_5\,(G_4 - 2X G_{4X})\,.
\eeq
%%%%%
In the context of the beyond Horndeski theory, the complexity of the field equations increases further,
and the integrability of the system of equations is easily lost. However, by employing a number of auxiliary functions,
we managed to bring the complicated equations of motion to a very simple form, namely \cite{BCKL}
%%%%%%
\begin{align} 
&X' \A = 2\left( \frac{h'}{h}- \frac{f'}{f}\right)\B\,,\nonumber\\
&\frac{h' f}{2 h} \A= G_{2 X} r^2 + 2 G_{4X} -2 r f \phi' G_{3X}-2 f Z_X\,, \\
&2 f \frac{h'}{h} \B=- G_2 r^2 -2 G_4 -2 f Z\,,\nonumber
\end{align}
where $\A$, $\B$ and $Z$ are functions of $\{G_i, F_i, X\}$ given by the expressions
\begin{align} 
&\A = 4r Z_X + \phi'\left[r^2 G_{3X} +G_{5X} (1-3f) -2X f G_{5XX} + 12 fX (5F_5 +2X F_{5X})\right],\nonumber\\
&\B=r Z -f \phi' X G_{5X} + 12 f \phi' X^2 F_5\,,\\
& Z =2X G_{4X}-G_4 + 4X^2 F_4 \,, \nonumber
\end{align}
and $(h,f)$ are the two metric functions of the static, spherically-symmetric line-element (\ref{metric_non_homo}).  

In the context of the parity-symmetric theory, i.e. with $G_3=G_5=F_5=0$, the above set of field equations is easily
rendered integrable and may lead to a large number of analytic solutions.  For instance, for the general class of theories
where 
%%%%%%%%%
\beq
G_2=-2\Lambda - \alpha X + \delta X^m\,, \qquad G_4= \zeta + \beta X^n\,,
\eeq
where $(\Lambda, \alpha, \delta, \zeta, \beta)$ are constant parameters of the theory and $(m,n)$ real, integer or
rational numbers,  and $Z$ chosen to be a constant, a large number of analytical homogeneous solutions
with $f(r)=h(r)$ were derived \cite{BCKL}. These solutions were characterized at the small-$r$ regime by zero,
one or two horizons, depending on the values of the parameters of the theory, while at large distances they
assumed an (Anti-)de Sitter-Reissner-Nordstrom-type of behaviour of the form 
%%%%%%%
\beq
h(r)=1- \frac{\Lambda_{eff} r^2}{3} -\frac{2 M}{r} + \frac{Q^2}{r^2}\,.
\eeq
%%%%%%%
In the expression above, $M$ is the ADM mass of the solution, $\Lambda_{eff}$ the effective cosmological constant and
$Q$ a tidal charge.  The first parameter is an arbitrary integration constant while the latter two are 
determined by the coupling parameters of the theory. All solutions are characterized by a non-trivial scalar
field, which at large distances behaves as $\phi' \sim 1/h(r)$ while it diverges at the (outer) horizon, when the latter
exists; however, $X$ remains everywhere finite as also does the energy-momentum tensor of the theory.
By making appropriate, alternative choices for the coupling functions of the theory, non-homogeneous solutions,
with $f \neq h$ but with the same asymptotic behaviour, were also found.

In the case of no parity symmetry, when the functions $(G_3, G_5, F_5)$ in principle do not vanish, the 
set of equations becomes significantly more difficult to solve. Nevertheless, even in this case, our formalism
allows for the integration of this set, under convenient choices for the coupling and auxiliary functions of
the theory, and new black-hole solutions, which generalise the solution of \cite{Lu_Pang}, can be determined,
albeit in a non-explicit form over the entire radial regime \cite{BCKL}. Additional black-hole solutions determined
in the context of Horndeski or beyond Horndeski theory can be found in \cite{Babichev:2016}-\cite{Babichev:2023}.

In the context of beyond Horndeski theory, wormhole solutions may also be found. However, instead of solving
the set of field equations for the desired type of solution, one could follow an alternative technique \cite{BCK}
\cite{Chatzifotis}
and apply an appropriate {\it disformal transformation}  \cite{Zumala}\cite{Crisostomi:2016}\cite{BenAchour_degen} to a
known solution denoted by $(\bar g_{\mu\nu}, \bar \phi)$. A disformal transformation has the general form 
\beq
g_{\mu \nu}=\bar{g}_{\mu \nu}- D(\bar{X})\; \nabla_\mu \phi\,\nabla_\nu \phi\,, \qquad \phi = \bar\phi\,,
\eeq
%%%%%
where $D(X)$ is an arbitrary function which characterizes the disformal transformation. According to the
above, the components of the metric tensor get transformed in a way that depends on the configuration of
the scalar field while the latter remains unchanged. In this way, one could start from a known solution
$(\bar g_{\mu\nu}, \bar \phi)$ and obtain a new one $(g_{\mu\nu}, \phi)$. When the ``seed
solution" $(\bar g_{\mu\nu}, \bar \phi)$ is a solution of Horndeski theory, the disformally transformed solution
$(g_{\mu\nu}, \phi)$ is a solution of the {\it beyond Horndeski} theory \cite{Crisostomi:2016}.

\begin{figure}[t]
\hspace*{2.5cm}
\mbox{\includegraphics[height=1.7in]{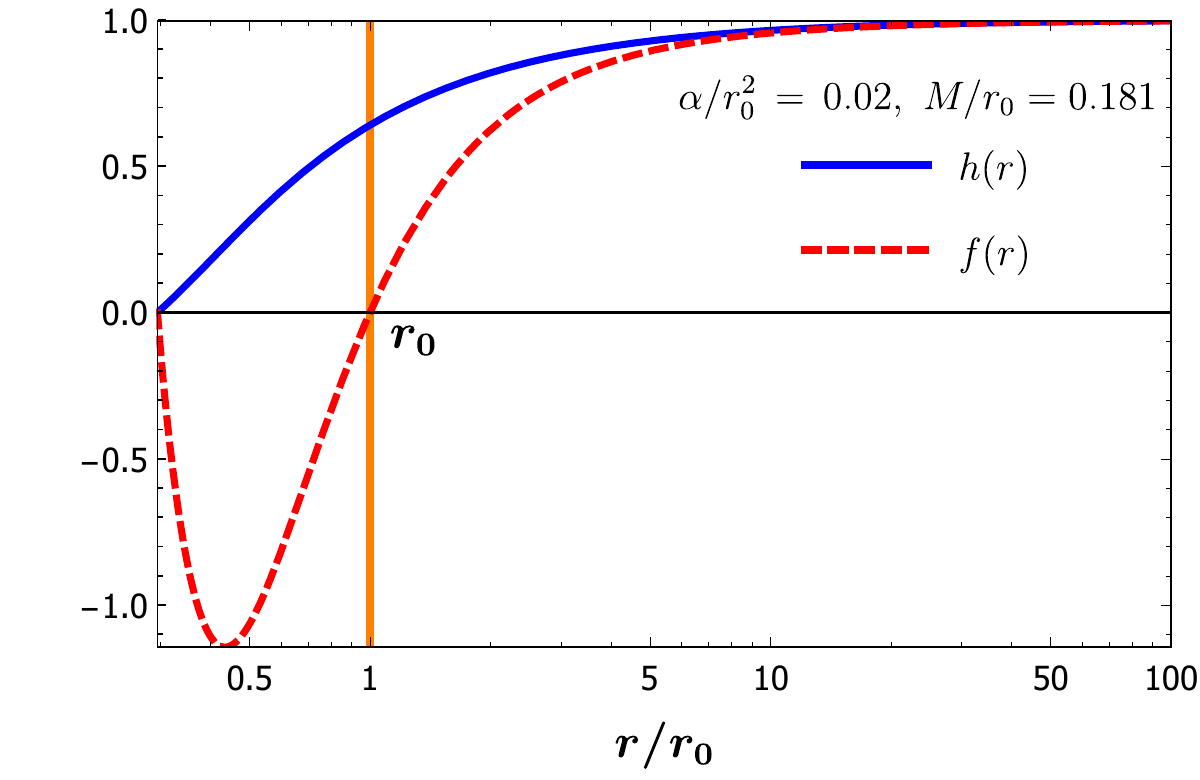}}
\caption{The two metric functions after the disformal transformation \cite{BCK}.}
\label{fig:throat}
\end{figure}

In our analysis \cite{BCK}, we chose to consider as our ``seed" solution $(\bar g_{\mu\nu}, \bar \phi)$
the static, spherically-symmetric solution (\ref{Lu_Pang}). Since  $\phi=\phi(r)$, the disformal transformation
will result in the relations
\beq
h=\bar h\,, \qquad f=\frac{\bar{f}}{1+2 D \bar{X}} \equiv \bar{f}\,W(\bar X)\,,
\eeq
where, for convenience, we have defined a new arbitrary function $W(\bar X)$. There is clearly an arbitrarily
large number of choices for the function $D(\bar X)$ or $W(\bar X)$. Its form will be dictated by the desired
properties of the new metric tensor. For instance, it is clear that $W(\bar X)$ should satisfy the following
two constraints
\beq
W(\bar X) \geq 0\,, \qquad \lim_{r \rightarrow \infty} W(\bar X)=1\,,
\eeq
in order to preserve the signature and asymptotic behaviour of $\bar g_{\mu\nu}$.  In addition, in order to
obtain a spacetime with a throat, we allowed the existence of a root in the expression of $W(\bar X)$ at a
distance $r_0 > r_H$, where $r_H$ is the event horizon of the ``seed solution". The presence of this root would
result into the vanishing of $g^{rr}$ at this point but not of $g_{tt}$, which will remain a constant. This
is indeed the typical behaviour of the metric tensor close to a wormhole according to the discussion in
Section 3.2. For example, we could choose \cite{BCK}
\beq
W(\bar X)=1-\frac{r_0}{\lambda}\,\sqrt{-2\bar X} =1-\frac{r_0}{\lambda  r}\left( 1-\sqrt{h}  \right)\,,
\eeq
where $\lambda$ is a scale parameter of the solution. The root $r_0$, or the radius of the wormhole, is located at 
\beq
r_0=\frac{M\pm \sqrt{M^2- \alpha \lambda^3\,(2-\lambda)^3}}{\lambda(2-\lambda)}\,,
\eeq
and is determined by the mass $M$ of the original black-hole solution, the coupling parameter $\alpha$
and the scalar parameter $\lambda$. The profiles of the two metric functions $h(r)$ and $f(r)$ after the
disformal transformation are depicted in Fig. \ref{fig:throat}. 

%%%%%%%%%%%%%%%%%%%
\begin{figure}[t!]
%\hspace*{6.5cm}
\mbox{\includegraphics[height=1.51in]{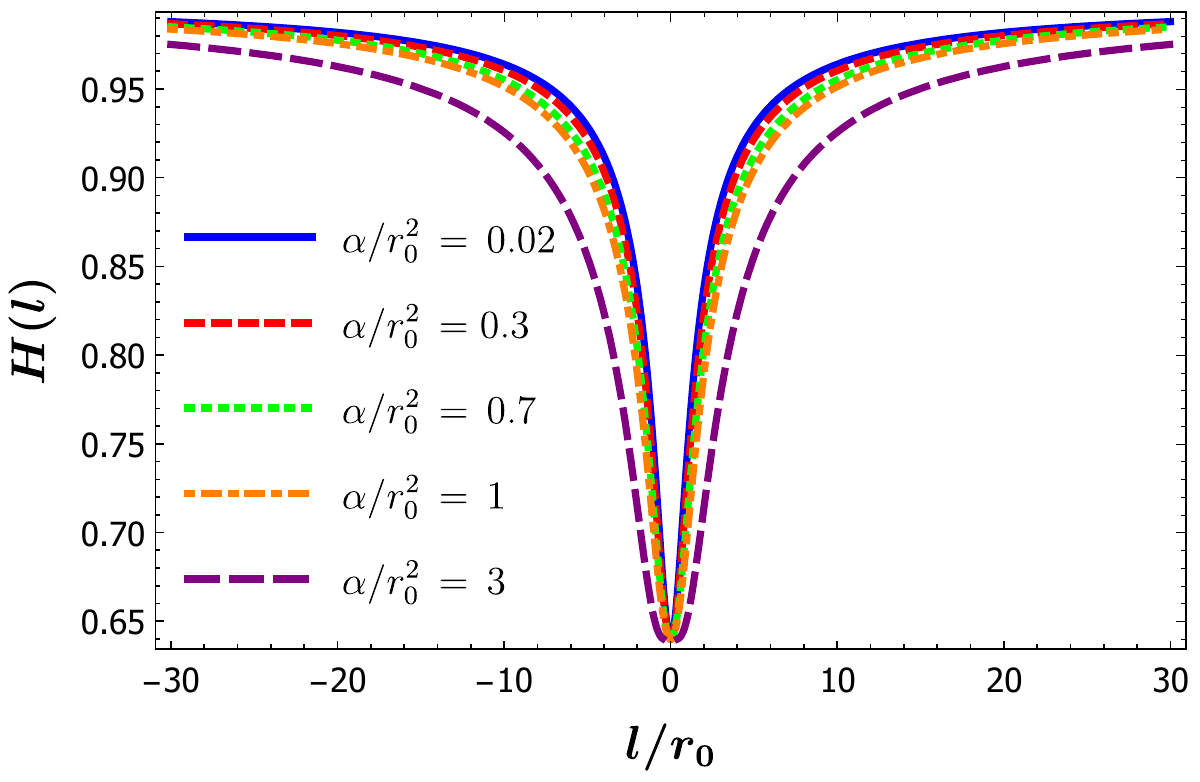}} \hspace*{0.5cm}
\includegraphics[height=1.51in]{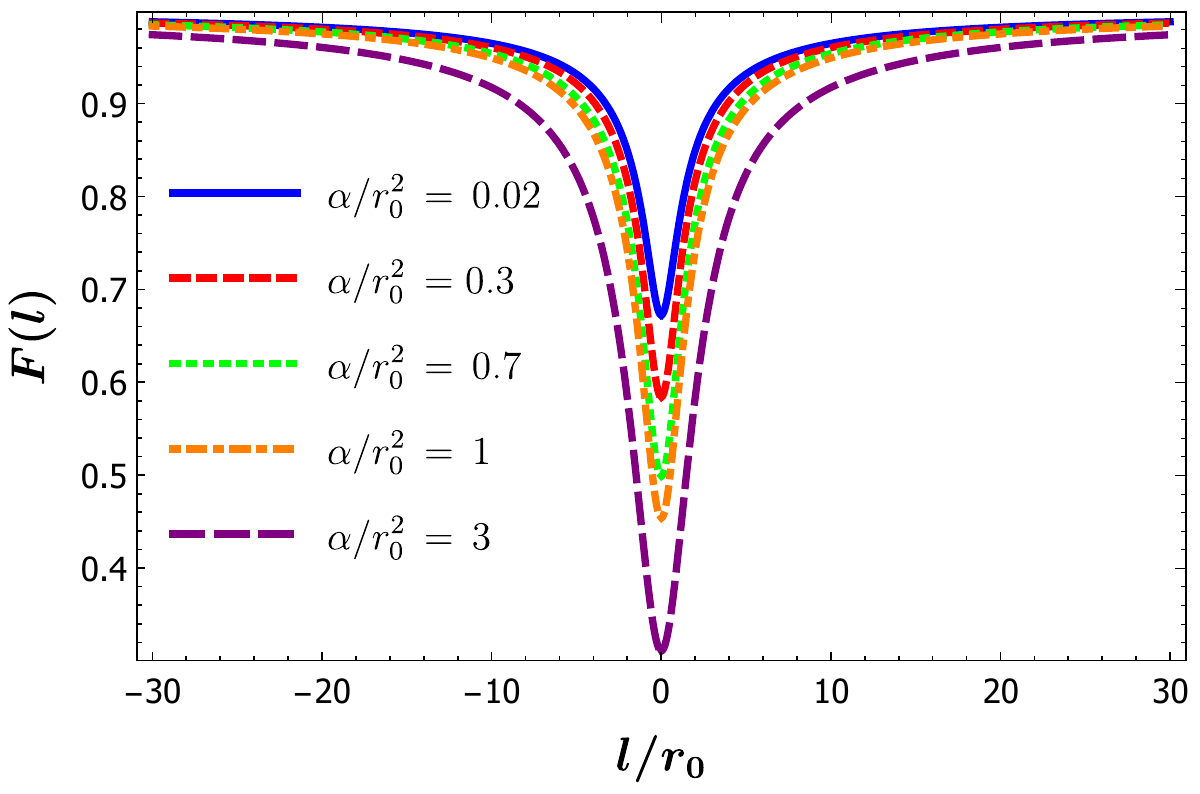} % requires the graphicx package
\caption{The two metric functions after the change of variable $r \rightarrow l$ \cite{BCK}.}
\label{fig:new_metric}
\end{figure}

Due to the vanishing  of the metric function $f(r)$ at the throat, we focus on the causal part of the
spacetime with $r_0 \leq r < \infty$. The complete geometry is revealed by setting $r^2= l^2 + r_0^2$, 
in which case the line-element reads
\beq
ds^2=-H(l) \,dt^2+\frac{dl^2}{F(l)}+(l^2+r_0^2)\,d\Omega^2\,,
\eeq
where
\beq
H(l)=h(r(l))\,, \qquad F(l)=\frac{f(r(l))\,(l^2 + r_0^2)}{l^2}\,.
\eeq
The throat of the wormhole is now located at $l=0$ and the two asymptotic regimes are reached when
$l \rightarrow \pm \infty$. The two metric functions $H(l)$ and $F(l)$ are everywhere regular and they
both assume constant values at the throat. Their profiles are depicted in Fig. \ref{fig:new_metric}. We
observe that both metric functions are symmetric under the change $l \rightarrow -l$ and that their
first derivatives vanish at the throat (all curves in Fig. \ref{fig:new_metric} become horizontal there).
As a result, no cusp points appear at the throat and the two parts of the wormhole (the positive and
negative $l$-regimes) are smoothly connected. The wormhole is therefore traversable without the
need of any distribution of additional matter, exotic or not.

%%%%%%%%%%%%%%%%%%%

\begin{figure}[htbp]
%\hspace*{6.5cm}
\mbox{\includegraphics[height=1.5in]{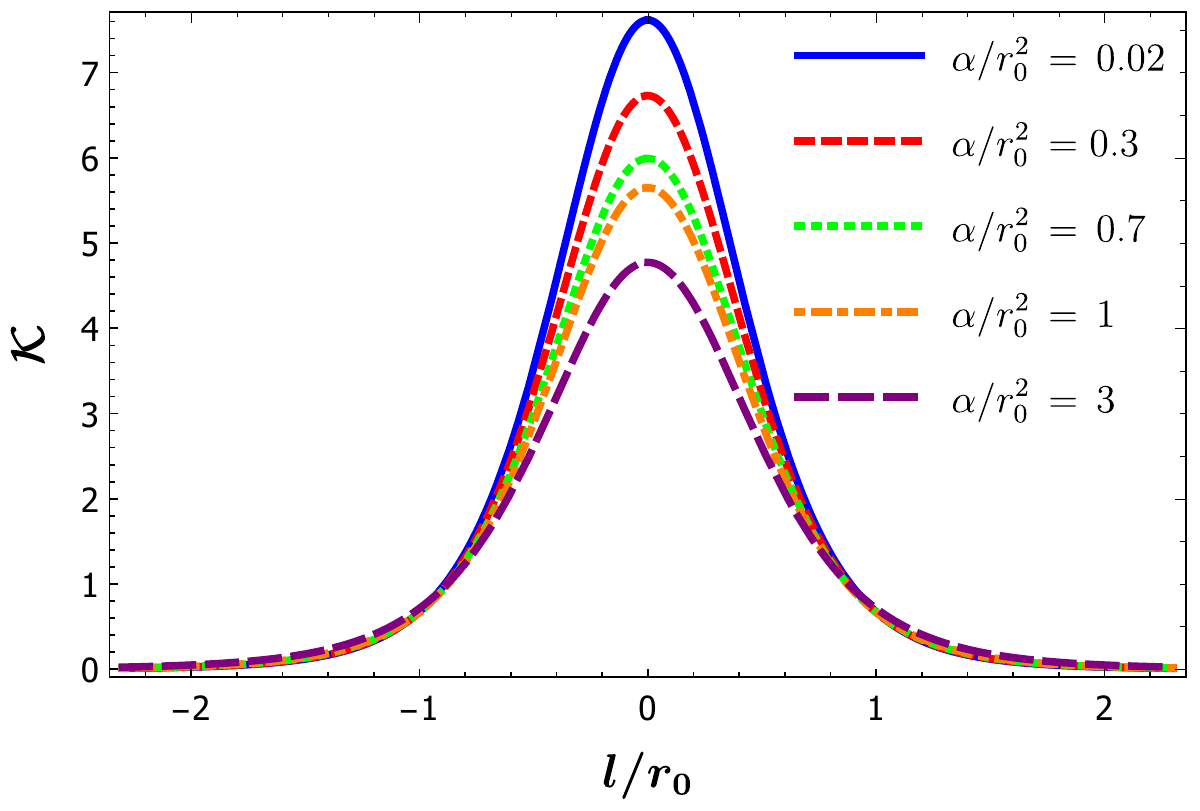}} \hspace*{0.7cm}
\includegraphics[height=1.5in]{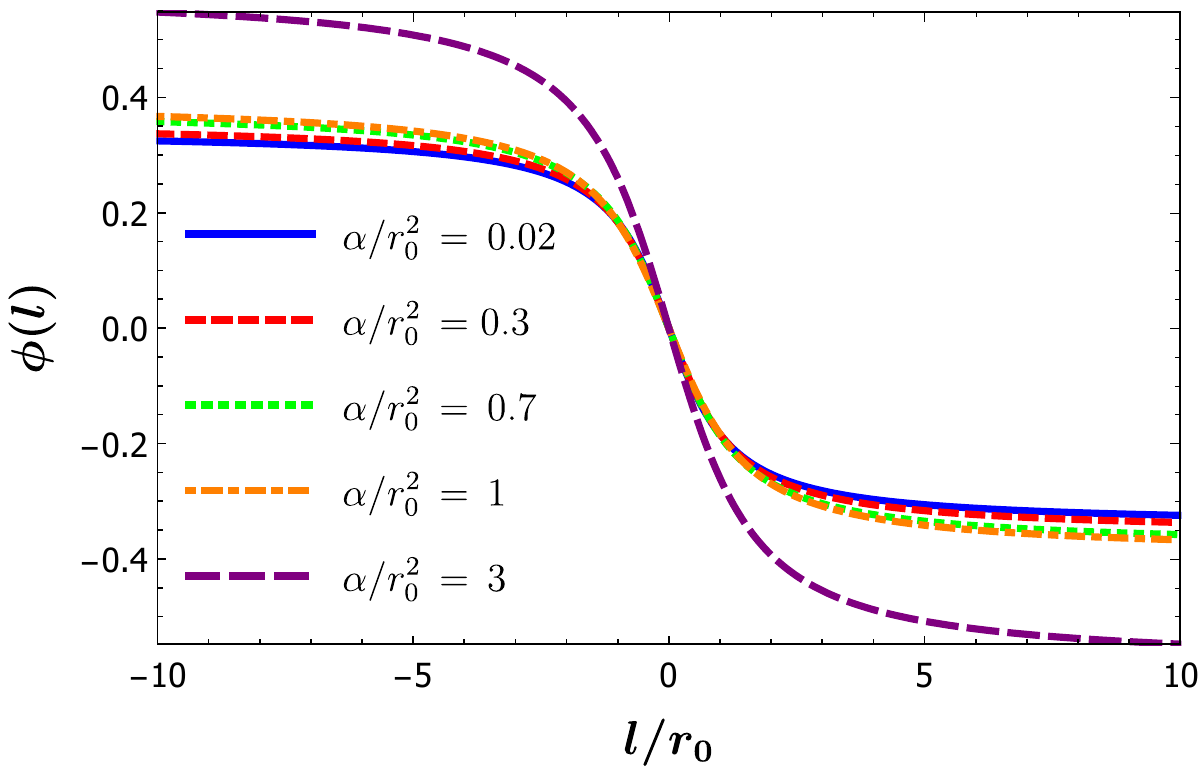} % requires the graphicx package
\caption{The Kretchmann scalar ${\cal K}$ (left plot), and the profile of the scalar field (right plot) \cite{BCK}. }
\label{fig:K_phi}
\end{figure}

The spacetime is everywhere regular and smooth, and this is evident in the form of the Kretchmann scalar
${\cal K}=R^{\mu\nu\rho\sigma} R_{\mu\nu\rho\sigma}$ presented on the left plot of Fig. \ref{fig:K_phi}.
We also observe that the non-vanishing curvature is restricted in a narrow region $-2r_0 \leq l \leq 2r_0$
thus our wormholes may comprise ultra-compact objects. One could construct the embedding diagram
of the solution by following a procedure similar to that of Section 3.2. Although the details differ, the
embedding diagram presents exactly the same form as the one shown in Fig. \ref{fig:embed}.
The profile of the scalar field is depicted on the right plot; this in fact asymmetric under the change
$l \rightarrow -l$, but is again regular over the entire regime and presents no discontinuities.

%%%%%%%%%%%%%%%%%%%

\begin{figure}[t]
%\hspace*{6.5cm}
\mbox{\includegraphics[height=1.5in]{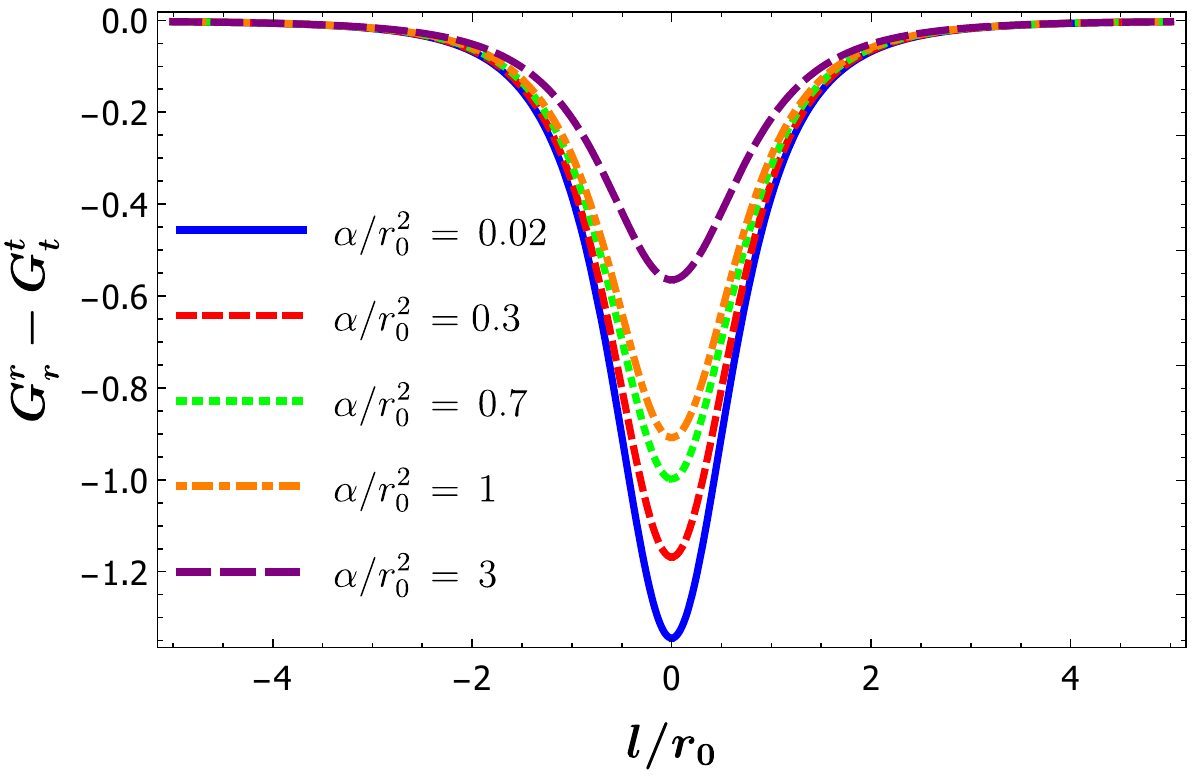}} \hspace*{0.7cm}
\includegraphics[height=1.5in]{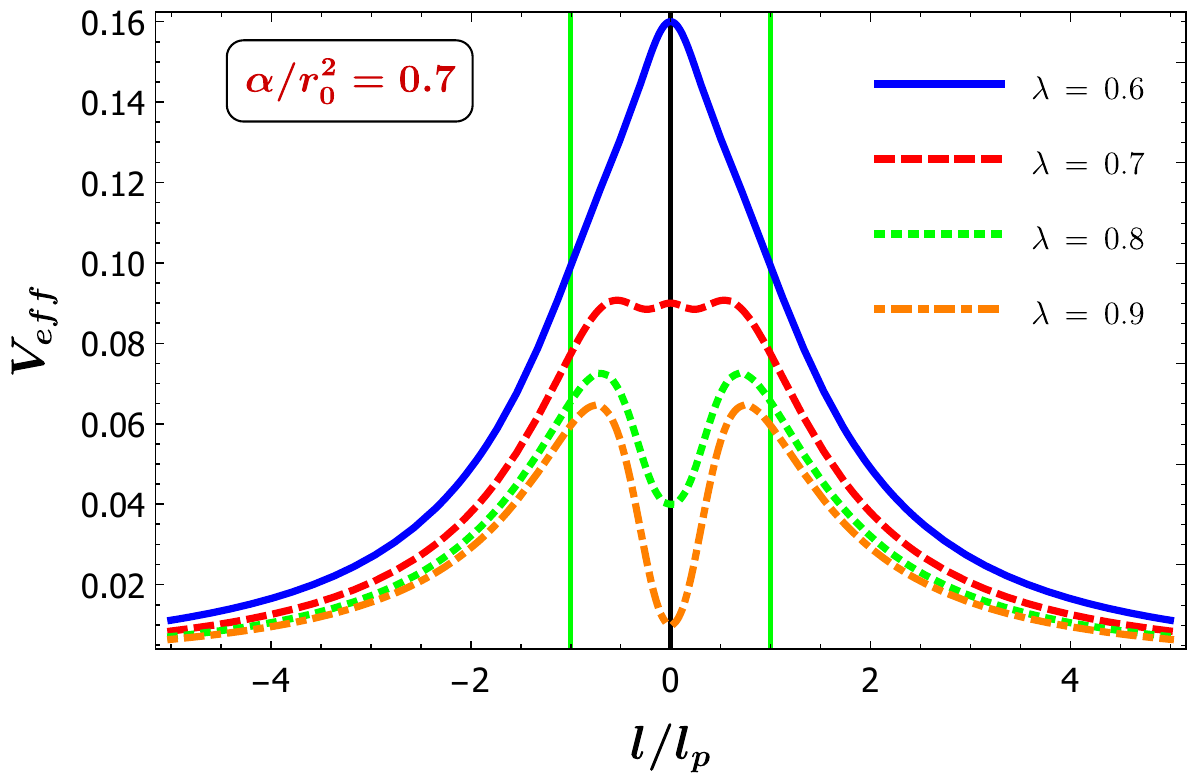} % requires the graphicx package
\caption{The violation of the NEC (left plot) and the effective potential of photons propagating in the
wormhole background (right plot) \cite{BCK}.}
 \label{fig:NEC}
\end{figure}

One may wonder whether the emergence of these wormhole solutions must again be supported by the
violation of the energy conditions. If we focus again on the Null Energy Condition, namely 
$T_{\mu\nu} n^\mu n^\nu \geq 0$, we arrive at the result
%%%%%
\beq
8\pi G (T^r_{\,\,r}-T^t_{\,\,t}) =G^r_{\,\,r}-G^t_{\,\,t}= - \frac{f'(r_0)}{r_0}\,.
\eeq
%%%%%%%
But one may easily see that the condition $f'(r_0) <0$ is the flaring-out condition for these wormholes
and thus must be always respected. As a result, the NEC is again violated, as one may see also on the left plot
of Fig. \ref{fig:NEC}. But this violation is again caused not by the presence of any exotic matter but by the
various non-minimal couplings of the scalar field to gravity in this theory. What is also important is
that the energy-density $\rho$ of the theory for most of the solutions is everywhere positive which renders
our solutions more physically interesting and realistic.

%%%%%%%%%%%%%%%%%%%
%%%%%%%%%%%%%%%%%%%

Turning finally to observable phenomena that one could associate with our wormhole solutions, we address 
again the geodesics of null particles as an indicative example. These are given by the equation
%%%%%
\beq
W \dot r^2 + \frac{h}{r^2} = \frac{E^2}{L^2}\,.
\eeq
Studying the extremal points of the photon's effective potential $V_{eff}=h/r^2$, we find that these are
given by the 3rd-order polynomial
\beq
r^3 - 9 M^2 r + 8 \alpha M=0\,.
\eeq
As expected, in the limit $\alpha \rightarrow 0$, we obtain two extremal points at $r=0$ and $r=3M$, with
the first one being stable and the second unstable. As $\alpha$ increases, we find up to 3 extremal points,
one at the throat and two more at larger radii. In this case, the light ring around the throat and the more
distant one are stable while the intermediate one is unstable; all light rings lie at distances smaller than
$3M$, which is the characteristic light-ring radius of the Schwarzschild spacetime. The effective potential
for null particles is depicted on the right plot of Fig. \ref{fig:NEC}.

%%%%%%%%%%%%%%%%  SLIDE 18 %%%%%%%%%%%%%%%%%%%%%%

\section{Conclusions }

In the quest of new black-hole solutions, traversable wormholes and regular particle-like solutions,
one is forced to move beyond General Relativity and to consider generalised theories of gravity. 
Scalar-tensor theories is the simplest extension of GR, however, one must consider non-minimally
coupled scalar fields in order for physically-interesting solutions to emerge. The Einstein-scalar-Gauss-Bonnet
theory is an indicative example of such a theory that contains a single scalar degree of freedom 
coupled to the quadratic, gravitational Gauss-Bonnet term, a term which is bound to lead
to important modifications from GR in the strong-gravity regime. It was in the context of this theory
that we sought for new solutions describing compact objects and being characterized by a non-trivial
scalar field. 

We demonstrated that the EsGB theory admits a variety of solutions. Scala- rised black-hole solutions
were discovered first for particular forms of the coupling function between the scalar field and the
GB term. In the recent years, it was demonstrated that scalarised, black-hole solutions with a regular
horizon and a Minkowski or Anti-de Sitter behaviour at large distances always emerge, independently
of the form of the coupling function, provided
that appropriate boundary conditions are imposed. These black-hole solutions evade all known
forms of the scalar no-hair theorem as the presence of the GB term causes the violation of the
theorem's requirements.

Similar analyses for the determination of wormhole solutions soon followed based on the common
knowledge that the GB term violates the energy conditions, a requirement for the emergence of 
a traversable wormhole. Indeed, a plethora of wormhole solutions were found for various forms of
the coupling function of the scalar field to the GB term. In order to avoid a spacetime singularity 
lurking somewhere behind the throat, a cut \& paste technique was applied which smoothly connected
two regular regimes thus creating a traversable wormhole with a regular scalar field. The study of
particle geodesics in this background showed that trajectories starting from one side of the wormhole,
crossing the throat, visiting the other side of the wormhole and then returning back to the departure
point was a common behaviour. 

Gravitational, particle-like solutions were also studied in the context of the EsGB theory causing
perhaps no surprise when such solutions readily emerged for various forms of the coupling
function. A singularity seemed to plague the expression of the scalar field, however, it was found
that this has no physical consequence for the spacetime curvature, energy-momentum tensor
or particle trajectories, thus rendering this singularity a harmless, ``Coulomb-type'' singularity.
These regular, particle-like solutions presented a bubble-type distribution of matter with a negative
value of energy density at the center, a positive value at the peak of the distribution located very
close to the origin and a fast decreasing profile - the latter features render our solutions as
ultra-compact objects. The study of null particles in the background of the more compact solutions 
revealed the existence of a pair of light-rings, an inner stable one and and an outer unstable.
The study of a scalar field propagating in the same spacetime led to the result that a series
of echoes should characterize the wave signal at infinity, a characteristic observable associated
to our particle-like solutions. 

\begin{figure}[t]
\hspace*{2.5cm}
\includegraphics[height=1.7in]{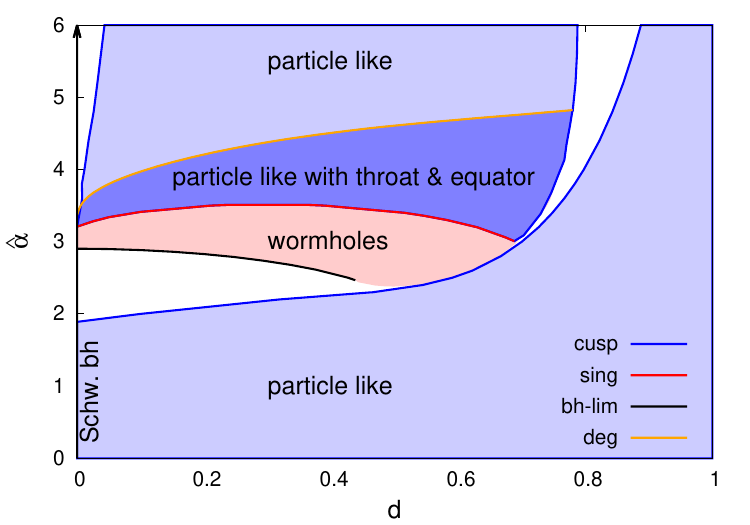} % requires the graphicx package
\caption{The different types of solutions emerging in the context of the Einstein-scalar-GB theory
\cite{KKK_part1}\cite{KKK_part2}.}
\label{fig:phase-space}
\end{figure}

The different types of solutions emerging in the context of the EsGB theory are depicted in the
domain-of-existence plot of Fig. \ref{fig:phase-space}. In there, we may see the line of scalarised black-hole
solutions, a one-parameter family of solutions due to the no-hair theorem which acts as a boundary
of the wormhole territory. The latter solutions may be characterised by a single throat or by a pair
of throats with an equator in between. In the same plot, we also notice different families of
particle-like solutions, characterised by an increasing number of nodes of the scalar field as the coupling constant
of the theory increases, too. We also observe that some white, uncharted yet, areas still remain in
the domain of existence which creates perhaps expectations for future discoveries!

%\vskip -0.2cm
%%%%%%%%%%%%%%%%%%%
%
%%%%%%%%%%%%%%%%%

The transition from General Relativity to the Einstein-scalar-Gauss-Bonnet theory resulted in the
discovery of a wealth of solutions which, in the former theory, were forbidden. The EsGB theory 
itself is a subclass of Horndeski theory, which is a much more general scalar-tensor theory
characterized by four coupling functions. Although more complicated, the set of field equations
can be rendered integrable upon appropriate choices for the four coupling functions. The theory
can be further generalised to the beyond-Horndeski theory via the addition of two additional terms in the
Lagrangian. In the context of this latter theory, we re-examined the integrability of the set of field equations, 
formulated these in a particularly convenient form, and investigated the existence of analytical
solutions describing black holes. In the context of the parity-symmetric sector of the theory,
we analytically determined a large number of black-hole solutions with an (Anti-)de Sitter-Reissner-Nordstrom
asymptotic behaviour at large distances. In the case of parity-symmetry breaking, the determination
of analytic solutions in a closed form is much more difficult. 

Turning to the family of wormhole solutions, we chose to determine these not by analytically solving the
field equations  but by applying a disformal transformation to a known, ``seed'', solution of the theory. 
That was made possible by conveniently choosing the function of the disformal transformation. 
The resulting wormhole solutions were shown to be generically regular by construction, localised
in a very narrow range of the radial coordinate -- a feature which renders our solutions as ultra-compact
objects -- and to be characterised by a non-trivial scalar field. Studying the particle trajectories in the
wormhole spacetime, distinct observable signatures were determined such as the emergence of 
multiple photon rings, some of them stable and all located at radii smaller than the characteristic
distance of $3M$ of the Schwarzschild solution.

Due to the larger arbitrariness of the Horndeski and beyond-Horndeski theories, encoded in the
forms of the four and six, respectively, undetermined a priori coupling functions, the phase-space
of the solutions of the theory has not yet fully been studied. Specific sectors have been studied for
particular classes of solutions but the study of the complete theory as well as the charting of the
types of solutions that the theory harbors is still lacking. Judging by the variety of solutions that
have so far been found and by the wealth of solutions arising in the context of the EsGB theory,
which is only a subclass of the theory, we expect a great many discoveries -- even surprises --
to arise when the complete Horndeski and beyond Horndeski theories are fully investigated.

{\bf Acknowledgements.} I am grateful to my collaborators Georgios Antoniou, Athanasios Bakopoulos,
Christos Charmousis, Nicolas Lecoeur, Burkhard Kleihaus, Jutta Kunz and Nikolaos Pappas for our
enjoyable and fruitful collaboration. I am also deeply thankful to the organisers of the 11th Aegean Summer
School for their kind invitation and for giving me the opportunity to present our results.


\end{document}